\documentclass[aps,showpacs,nofootinbib,superscriptaddress,twocolumn,floatfix]{revtex4-1}

\usepackage{graphicx}
\usepackage{amssymb}
\usepackage{amsmath}
\usepackage{bm}
\usepackage{slashed}
\usepackage{datetime}
\usepackage{mciteplus}

\newcommand{\open}{\sphericalangle}
\newcommand{\sT}{\scriptstyle T}

\begin{document}
\allowdisplaybreaks[2]

\title{First extraction of Interference Fragmentation Functions from $e^+ e^-$ data}

\author{Aurore Courtoy}
\email{aurore.courtoy@ulg.ac.be}
\affiliation{INFN Sezione di Pavia, via Bassi 6, I-27100 Pavia, Italy}
\affiliation{IFPA, AGO Department, Universit\'e de Li\`ege, Belgium}

\author{Alessandro Bacchetta}
\email{alessandro.bacchetta@unipv.it}
\affiliation{Dipartimento di Fisica Nucleare e Teorica, Universit\`a di Pavia}
\affiliation{INFN Sezione di Pavia, via Bassi 6, I-27100 Pavia, Italy}

\author{Marco Radici}
\email{marco.radici@pv.infn.it}
\affiliation{INFN Sezione di Pavia, via Bassi 6, I-27100 Pavia, Italy}

\author{Andrea Bianconi}
\email{andrea.bianconi@bs.infn.it}
\affiliation{Dipartimento di Chimica e Fisica per l'Ingegneria e per i Materiali, Universit\`a di Brescia, I-25123 Brescia, Italy}
\affiliation{INFN Sezione di Pavia, via Bassi 6, I-27100 Pavia, Italy}

\begin{abstract}
We report on the first extraction of interference fragmentation functions from the semi-inclusive production of two hadron pairs in back-to-back jets in $e^+ e^-$ annihilation. A nonzero asymmetry in the correlation of azimuthal orientations of opposite $\pi^+ \pi^-$ pairs is related to the transverse polarization of fragmenting quarks through a significant polarized dihadron fragmentation function. Extraction of the latter requires the knowledge of its unpolarized counterpart, the probability density for a quark to fragment in a $\pi^+ \pi^-$ pair. Since data for the unpolarized cross section are missing, we extract the unpolarized dihadron fragmentation function from a Monte Carlo simulation of the cross section.  
\end{abstract}



\maketitle


\section{Introduction}
\label{sec:intro}

In the hadronization process, there is a nonvanishing probability that at a hard scale $Q^2$ a highly virtual parton fragments into two hadrons inside the same jet, carrying fractional energies $z_1$ and $z_2$, plus other unobserved fragments. This nonperturbative mechanism can be encoded in the so-called dihadron fragmentation functions (DiFFs) of the form $D(z_1, z_2; Q^2)$. The interest in two-particle correlations 
in $e^+ e^-$ processes was first pointed out in Ref.~\cite{Walsh:1974bj}. DiFFs were introduced for the first time in the context of jet calculus~\cite{Konishi:1978yx}, and they are needed to cancel all collinear singularities when the semi-inclusive production of two hadrons from $e^+ e^-$ annihilations is considered at next-to-leading order in the strong coupling constant~\cite{deFlorian:2003cg} (NLO). 

Experimental information on two hadron production is often delivered in terms of a distribution in the invariant mass $M_h$ of the hadron pair~\cite{Acton:1992sa,Abreu:1992xx,Buskulic:1995gm}. Therefore, it is convenient to describe the process with ``extended" DiFFs of the form 
$D(z_1, z_2, M_h; Q^2)$, in analogy to what  is done for fracture functions~\cite{Grazzini:1997ih}. If 
$M_h^2 \approx Q^2$, DiFFs asymptotically transform into the combination of two single-hadron fragmentation functions~\cite{Zhou:2011ba}. If $M_h^2 \ll Q^2$, they represent a truly new  nonperturbative object. For polarized fragmentations, certain DiFFs emerge from the interference of amplitudes with the hadron pair being in two states with different relative angular 
momentum~\cite{Collins:1994ax,Jaffe:1998hf,Radici:2001na,Bacchetta:2006un}. Hence, in the literature they are addressed also as interference fragmentation functions (IFFs)~\cite{Jaffe:1998hf}. 
IFFs can be used in particular as analyzers of the polarization state of the fragmenting 
parton~\cite{Efremov:1992pe,Collins:1994kq,Artru:1995zu,Bianconi:1999cd}. 

The definition of DiFFs and a thorough study of their properties were presented in 
Refs.~\cite{Bianconi:1999cd,Bacchetta:2002ux} (up to leading twist) and in 
Ref.~\cite{Bacchetta:2003vn} (including subleading twist; see also Ref.~\cite{Gliske:2011}). At $M_h^2\ll Q^2$, DiFFs satisfy the same evolution equations as the single-hadron fragmentation 
functions~\cite{Ceccopieri:2007ip}, in contrast to what happens if DiFFs are integrated over $M_h^2$~\cite{deFlorian:2003cg}. They can be factorized and are assumed to be universal. In fact, they appear not only in $e^+ e^-$ annihilations~\cite{Boer:2003ya,Bacchetta:2008wb}, but also in hadron pair production in semi-inclusive deep-inelastic scattering 
(SIDIS)~\cite{Bacchetta:2002ux,Bacchetta:2008wb} and in hadronic 
collisions~\cite{Bacchetta:2004it}. 

The case of SIDIS production of $(\pi^+ \pi^-)$ pairs (or of any pair of distinguishable unpolarized hadrons) on transversely polarized protons is of particular interest. In fact, in the fragmentation 
$q^\uparrow \to (\pi^+ \pi^-) X$ a correlation occurs between the transverse polarization of the parton 
$q^\uparrow$ and the relative orbital angular momentum of the pair. Such nonperturbative effect is encoded in the chiral-odd DiFF 
$H_1^{\open\,q}$~\cite{Collins:1994ax,Jaffe:1998hf,Bianconi:1999cd}, which arises from the interference of fragmentation amplitudes $(\pi^+ \pi^-)_L$ with relative partial waves $L$ differing by $|\Delta L| = 1$~\cite{Jaffe:1998hf,Radici:2001na,Bacchetta:2002ux}. The $H_1^{\open\,q}$ appears in the factorized formula for the leading-twist SIDIS cross section in a simple product with the chiral-odd transversity distribution $h_1^q$~\cite{Bacchetta:2002ux}, the most elusive parton distribution,  needed to give a complete description of the collinear partonic spin structure of the nucleon (for a review, see Ref.~\cite{Barone:2003fy}). The same $H_1^{\open\,q}$ (and its antiquark partner) appears in the factorized formula for the leading-twist cross section for the process $e^+ e^- \to (\pi^+ \pi^-) (\pi^+ \pi^-) X$~\cite{Boer:2003ya,Bacchetta:2008wb}, where the transverse polarization of the elementary $q^\uparrow \overline{q}^\downarrow$ pair is correlated to the azimuthal orientation of the planes containing the momenta of the two pion 
pairs~\cite{Artru:1995zu,Boer:2003ya}. Thus, extracting the $h_1^q  H_1^{\open\,q}$ and 
$H_1^{\open\,q}  \overline{H}_1^{\open\, q}$ combinations through specific azimuthal asymmetries in SIDIS and $e^+ e^-$, respectively, offers a way to isolate the transversity $h_1^q$ with significant theoretical 
advantages~\cite{Radici:2001na,Bacchetta:2002ux,Boer:2011fh,Bacchetta:2011ip} with respect to the traditional strategy based on the Collins effect~\cite{Collins:1993kk}. 

The spin asymmetry in the SIDIS process $ep^\uparrow \to e' (\pi^+ \pi^-) X$ was measured by the HERMES collaboration~\cite{Airapetian:2008sk}; preliminary data are available also from the COMPASS collaboration~\cite{Wollny:2010}. Clear evidence for the required azimuthal asymmetry in the process $e^+ e^- \to (\pi^+ \pi^-) (\pi^+ \pi^-) X$ has been recently reported by the Belle 
collaboration~\cite{Vossen:2011fk}. A combined analysis of these data has led to the first extraction of transversity in the framework of collinear factorization using two-hadron inclusive 
measurements~\cite{Bacchetta:2011ip}. The results seem for the moment compatible with the only other available parametrization of $h_1^q$, which is based on the Collins effect in single-hadron 
production~\cite{Anselmino:2008jk}. However, more data are needed to strengthen the case, including also proton-proton collisions where preliminary results are available from the PHENIX collaboration~\cite{Yang:2009zzr}. 

All these analyses require a good knowledge of the dependence upon $z_1, z_2,$ and $M_h$ of the polarized DiFF $H_1^{\open\,q}$, as well as of its polarization-averaged partner $D_1^q$. In this paper, we study the full dependence of  both DiFFs for the $u, d, s,$ and $c$ flavors. We start at a low hadronic scale $Q_0^2 = 1$ GeV$^2$ using a parametrization inspired by previous model calculations of DiFFs~\cite{Bianconi:1999uc,Bacchetta:2006un,Bacchetta:2008wb}. Then, we apply evolution equations to DiFFs using the {\tt HOPPET} code~\cite{Salam:2008qg}, suitably extended to include chiral-odd splitting functions. Finally, we fit the recent Belle data on azimuthal asymmetries in the orientation of $(\pi^+ \pi^-)$ pairs collected at $Q^2 = 100$ GeV$^2$ (close to the $\Upsilon (4S)$ resonance). In the absence of published data for the unpolarized cross section, we parametrize the $D_1^q$ by fitting the prediction of the {\tt PYTHIA} event generator~\cite{Sjostrand:2003wg} adapted to the Belle kinematics, since this code is known to give a good description of the total cross section~\cite{courtesyBELLE}. The information delivered by {\tt PYTHIA} is much richer than the asymmetry measurement. Consequently, the analysis for $D_1^q$ can be developed to a much deeper detail than what is possible for $H_1^{\open\,q}$. It is anyway useful to obtain a thorough knowledge of the unpolarized DiFF, even if based on ``virtual" data. In the future, we hope it will be possible to perform an analogous study on real data. 

The paper is organized as follows. In Sec.~\ref{sec:eqs}, we briefly recall the formalism for the 
$e^+ e^- \to (\pi^+ \pi^-) (\pi^+ \pi^-) X$ process. In Sec.~\ref{sec:D1}, we describe the steps leading to the extraction of $D_1^q$ from the Monte Carlo simulation. In Sec.~\ref{sec:H1}, we describe the extraction of $H_1^{\open\,q}$ from real data. Finally, in Sec.~\ref{sec:end} we discuss some outlooks for future improvements.


\begin{figure}[h]
\begin{center}
\includegraphics[width=8cm]{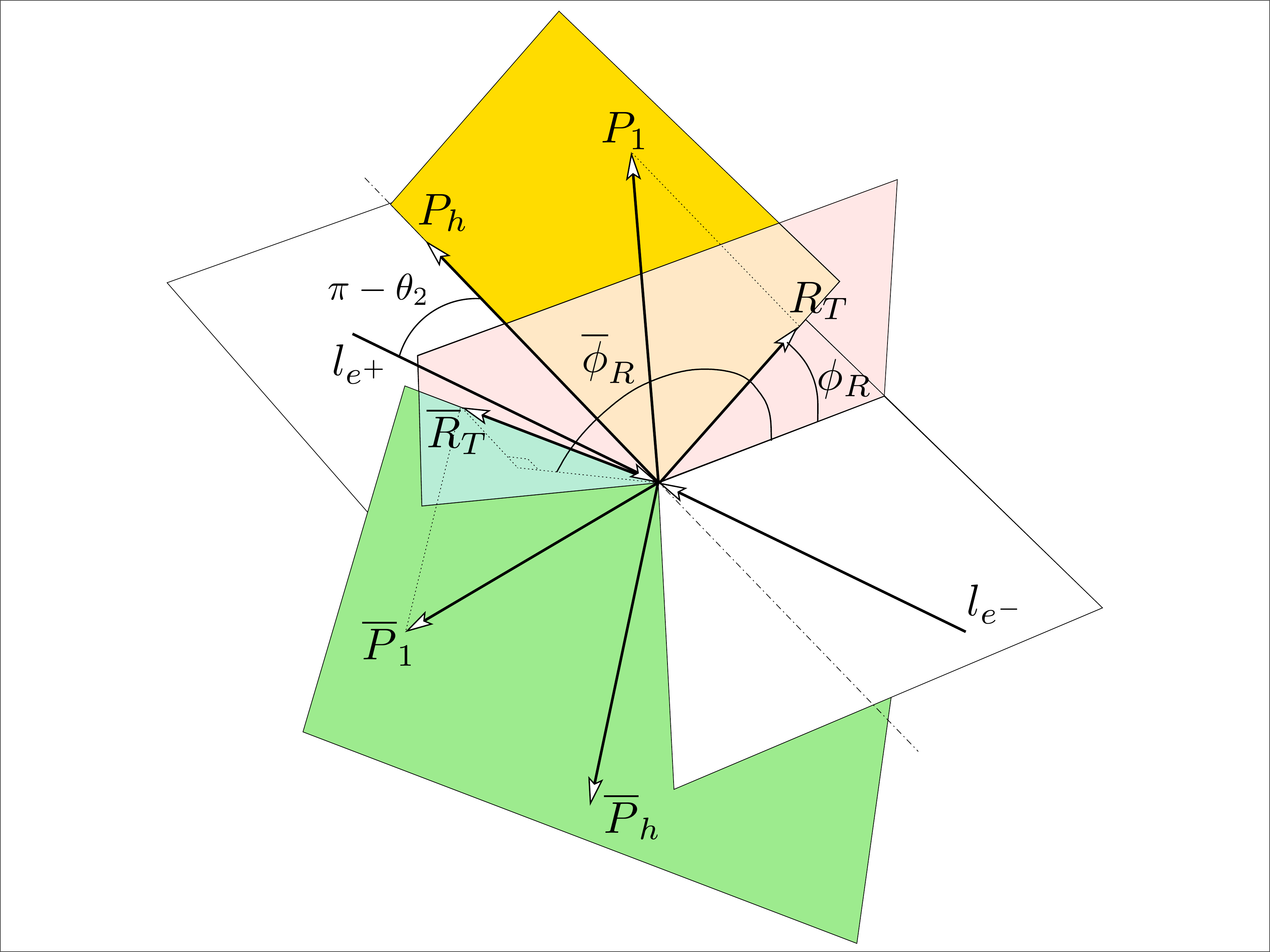}
\end{center}
\caption{Definition of the kinematics for the process 
$e^+ e^- \to (\pi^+ \pi^-)_{\mathrm{jet1}} (\pi^+ \pi^-)_{\mathrm{jet2}} X$.}
\label{fig:kin}
\end{figure}

\section{Formalism}
\label{sec:eqs}

We consider the process $e^+ e^- \to (\pi^+ \pi^-)_{\mathrm{jet1}} (\pi^+ \pi^-)_{\mathrm{jet2}} X$, depicted in Fig.~\ref{fig:kin}. An electron and a positron with momenta $l_{e^-}$ and $l_{e^+}$, respectively, annihilate producing a photon with time-like momentum transfer $q=l_{e^-}+l_{e^+}$, i.e. $q^2=Q^2 \geq 0$. A quark and an antiquark are then emitted, each one fragmenting into a residual jet and a $(\pi^+ \pi^-)$ pair with momenta and masses $P_1, M_1,$ and 
$P_2, M_2,$ respectively (for the pair in the antiquark jet, we use the notation 
$\overline{P}_1, \overline{M}_1,$ and $\overline{P}_2, \overline{M}_2,$ respectively, and similarly for all other observables pertaining the antiquark hemisphere). We introduce the pair total momentum $P_h = P_1+P_2$ and relative momentum $R=(P_1-P_2)/2$, and the pair invariant mass $M_h$ with $P_h^2 = M_h^2$. The two $(\pi^+ \pi^-)$ pairs belong to two back-to-back jets, from which $P_h\cdot \overline{P}_h \approx Q^2$. Using the standard notations for the light-cone components of a 4-vector, we define the following light-cone fractions 
\begin{align}
z =\frac{P_h^-}{q^-} = z_1+z_2  \qquad &\zeta = 2\frac{R^-}{P_h^-}=\frac{z_1-z_2}{z}  \nonumber \\
\overline{z}=\frac{\overline{P}_h^+}{q^+}=\overline{z}_1+\overline{z}_2 \qquad 
 &\overline{\zeta} = 2\frac{\overline{R}^+}{\overline{P}_h^+}= 
 \frac{\overline{z}_1-\overline{z}_2}{\overline{z}} \; . 
\label{e:invariants}
\end{align}
The $z$ is the fraction of quark momentum carried by the pion pair, and $\zeta$ describes how the total momentum of the pair is split between the two pions~\cite{Bacchetta:2006un} (and similarly for 
$\overline{z}, \overline{\zeta},$ referred to the fragmenting antiquark). In Fig.~\ref{fig:kin}, we identify the lepton frame with the plane formed by the annihilation direction of $\bm{l}_{e^+}$ and the axis 
$\hat{\bm{z}} = - \bm{P}_h$, in analogy to the Trento conventions~\cite{Bacchetta:2004jz}. The relative angle is defined as 
$\theta_2 = \arccos (\bm{l_{e^+}}\cdot\bm{P}_h / (|\bm{l_{e^+}}|\,|\bm{P}_h|))$ and is related, in the lepton center-of-mass frame, to the invariant $y = P_h\cdot l_{e^-} / P_h \cdot q$ by 
$y = (1+\cos\theta_2)/2$. The azimuthal angles $\phi_R^{}$ and $\overline{\phi}_R^{}$ give the orientation of the planes containing the momenta of the pion pairs with respect to the lepton frame. They are defined by~\cite{Bacchetta:2008wb}
\begin{equation}
\begin{split}
\phi_R^{} &= 
\frac{(\bm{l}_{e^+}\times \bm{P}_h)\,\cdot \bm{R}_{\sT}}
     {|(\bm{l}_{e^+}\times \bm{P}_h)\,\cdot \bm{R}_{\sT}|} 
\arccos \left( 
        \frac{\bm{l}_{e^+}\times \bm{P}_h}{|\bm{l}_{e^+}\times \bm{P}_h|}
	\cdot 
	\frac{\bm{R}_{\sT}\times \bm{P}_h}{|\bm{R}_{\sT}\times \bm{P}_h|} 
	\right)
\\
\overline{\phi}_R^{} &= 
\frac{(\bm{l}_{e^+}\times \bm{P}_h)\,\cdot \overline{\bm{R}}_{\sT}}
     {|(\bm{l}_{e^+}\times \bm{P}_h)\,\cdot \overline{\bm{R}}_{\sT}|} 
\arccos \left( 
        \frac{(\bm{l}_{e^+}\times \bm{P}_h)}{|\bm{l}_{e^+}\times \bm{P}_h|}
        \cdot 
	\frac{(\overline{\bm{R}}_{\sT}\times \bm{P}_h)}
	     {|\overline{\bm{R}}_{\sT}\times \bm{P}_h|} 
	\right)  \, , 
\label{eq:az_angles}
\end{split}
\end{equation}
where $\bm{R}_{\sT}$ is the transverse component of $\bm{R}$ with respect to $\bm{P}_h$ (and similarly for $\overline{\bm{R}}_{\sT}$). The above framework corresponds in 
Ref.~\cite{Vossen:2011fk} to the frame where no thrust axis is used to define angles, and where all quantities are labelled by the subscript ``$R$". 

The previous definitions imply that~\cite{Bacchetta:2006un}
\begin{equation}
\frac{|\bm{R}|}{M_h} = \frac{1}{2}\,\sqrt{1-\frac{4m_\pi^2}{M_h^2}} \; .
\label{eq:|R|}
\end{equation}
Moreover, the light-cone fractions $\zeta, \overline{\zeta},$ can be rewritten 
as~\cite{Bacchetta:2006un}
\begin{align}
\zeta = 2 \frac{|\bm{R}|}{M_h}\, \cos\theta \qquad &\overline{\zeta} = 2 
\frac{|\overline{\bm{R}}|}{\overline{M}_h}\, \cos\overline{\theta} \; , 
\label{e:cos}
\end{align}
where $\theta$ describes the direction of $P_1$, in the center-of-mass frame of the pion pair, with respect to the direction of $P_h$ in the lepton frame (and similarly for $\overline{\theta}$ in the other hemisphere). From Eqs.~\eqref{e:invariants} and~\eqref{e:cos}, DiFFs depend directly on 
$z, \cos\theta,$ and they can be expanded in terms of Legendre polynomials of $\cos\theta$. We keep only the first two terms, which correspond to $L=0 \; (s)$ and $L=1 \; (p)$ relative partial waves of the pion pair~\cite{Bacchetta:2002ux}, since we assume that at low invariant mass the contribution from higher partial waves is negligible. 

Using the definitions and transformations above, we can start from Eq.~(30) of 
Ref.~\cite{Boer:2003ya} (see also Ref.~\cite{Bacchetta:2008wb}) and write the leading-twist unpolarized cross section for the production of two pion pairs (summing over everything else) as
\begin{equation}
\begin{split}
&\frac{d\sigma^0}{d\cos \theta_2\, dz\,d\cos\theta \, dM_h\,d\overline{z}\, d\cos\overline{\theta}\, 
d\overline{M}_h\,dQ^2} = \frac{3\pi \alpha^2}{2 Q^2} \\
&\quad \times \frac{1+\cos^2\theta_2}{4} \sum_q e_q^2\, D_1^q (z, M_h; Q^2)\, 
\overline{D}_1^q (\overline{z}, \overline{M}_h; Q^2) \; , 
\label{eq:dsig0}
\end{split}
\end{equation}
where the flavor sum is understood to run over quarks and antiquarks, and in the expansion of $D_1$ ($\overline{D}_1$) in Legendre polynomials of $\cos\theta$ ($\cos \overline{\theta}$) we have kept the first nonvanishing term after integrating in $d\cos\theta$ 
($d\cos \overline{\theta}$)~\cite{Bacchetta:2002ux}. 

The fully differential polarized part of the leading-twist cross section contains many terms (see 
Eq.~(19) in Ref.~\cite{Boer:2003ya}). But in the framework of collinear factorization, i.e. after integrating upon all transverse momenta but $\bm{R}_{\sT}$ and $\overline{\bm{R}}_{\sT}$, only one term survives beyond $d\sigma^0$. It is identified by its azimuthal dependence 
$\cos (\phi_R^{}+\overline{\phi}_R^{})$, which is responsible for the asymmetry in the relative position of the planes containing the momenta of the two pion pairs. Then, the integrated full cross section can be written as 
\begin{equation}
\begin{split}
&\frac{d\sigma}{d\cos \theta_2\, dz\,d\cos\theta \, dM_h\, d\phi_R^{}\, d\overline{z}\, 
d\cos\overline{\theta}\, d\overline{M}_h\, d\overline{\phi}_R^{}\, dQ^2} = \\
&\mbox{\hspace{1cm}} \frac{1}{4\pi^2}\, d\sigma^0\, \left( 1 + \cos (\phi_R^{}+\overline{\phi}_R^{})\, 
A \right) \; , 
\label{eq:dsig}
\end{split}
\end{equation}
where we define the socalled Artru--Collins azimuthal asymmetry (compare with Eq.~(21) in 
Ref.~\cite{Boer:2003ya} and Eq.~(11) in Ref.~\cite{Bacchetta:2008wb})
\begin{equation} 
\begin{split} 
&A(\cos\theta_2, z, \cos\theta, M_h, \overline{z}, \cos\overline{\theta}, \overline{M}_h, Q^2) 
=  \\
&\mbox{\hspace{1cm}} \frac{\sin^2 \theta_2}{1+\cos^2 \theta_2} \, \sin\theta \sin\overline{\theta}\, 
\frac{|\bm{R}|}{M_h} \, \frac{|\overline{\bm{R}}|}{\overline{M}_h} \\
&\mbox{\hspace{1cm}} \times \frac{\sum_q e_q^2 \, H_{1, sp}^{\open q}(z, M_h; Q^2)\,
          \overline{H}_{1, sp}^{\open q}(\overline{z}, \overline{M}_h; Q^2)}
      {\sum_q e_q^2\, D_1^q (z, M_h; Q^2) \, \overline{D}_1^q (\overline{z}, \overline{M}_h; Q^2)} \; .
\label{eq:ACasy}
\end{split} 
\end{equation} 
In the expression above, we have used the relation $\bm{R}_{\sT} = \bm{R} \sin\theta$ (and similarly for $\overline{\bm{R}}_{\sT}$). Again, in the expansion of DiFFs in Legendre polynomials of $\cos\theta$ ($\cos\overline{\theta}$) we have kept the first nonvanishing term after integrating in 
$d\cos\theta$ ($d\cos\overline{\theta}$)~\cite{Bacchetta:2002ux}. For the polarized part, this amounts to keep that component of $H_1^{\open q}$ corresponding to the interference between a pair in relative $s$ wave and the other one in relative $p$ wave, namely 
$H_{1, sp}^{\open q}$~\cite{Bacchetta:2008wb}. Note also that, at variance with 
Ref.~\cite{Boer:2003ya}, the azimuthally asymmetric term is not isolated by integrating over 
$\phi_R^{}$ and $\overline{\phi}_R^{}$, since the integration could not be complete in the experimental acceptance. Rather, it is extracted as the coefficient of the 
$\cos (\phi_R^{}+\overline{\phi}_R^{})$ modulation on top of the flat distribution produced by the unpolarized part. 

For our analysis, it is necessary to consider the unpolarized cross section $d\sigma^0$ also for the production of just one pion pair. From Eq.~\eqref{eq:dsig0}, we have
\begin{equation}
\begin{split}
&\int d\overline{z} \, d\overline{M}_h \, d\cos \overline{\theta} \quad d\sigma^0 
\Bigg\vert_{\overline{D}_1 = \delta (1-\overline{z})}  \\
&\qquad \equiv \frac{d\sigma^0}{d\cos \theta_2\, dz\,d\cos\theta \, dM_h\, dQ^2} \\
&\qquad = \frac{3\pi \alpha^2}{Q^2}  \frac{1+\cos^2\theta_2}{4}  \sum_q e_q^2\, 
D_1^q (z, M_h; Q^2) \; .
\label{eq:dsig0-2}
\end{split}
\end{equation}

Our strategy is the following. We start from a parametrization of DiFFs at the low hadronic scale 
$Q_0^2 = 1$ GeV$^2$ by taking inspiration from previous model 
analyses~\cite{Bianconi:1999uc,Bacchetta:2006un,Bacchetta:2008wb}. Then, we evolve DiFFs at leading order (LO) up to the Belle scale $Q^2 = 100$ GeV$^2$ by using the {\tt HOPPET} 
code~\cite{Salam:2008qg}, suitably extended to include chiral-odd splitting functions. In principle, the unpolarized $D_1$ should be extracted by global fits of the unpolarized cross section, in the same way as it is done for single-hadron fragmentation~\cite{deFlorian:2007aj}. Because no data are available yet, we extract it by fitting the single pair distribution simulated by a Monte Carlo event generator. Next, we fit the experimental data for the Artru--Collins asymmetry of 
Eq.~\eqref{eq:ACasy} and we extract $H_1^{\open}$ from this fit. In the following, we list some more details of our analysis and we discuss the final results.


\section{Extraction of $D_1$ from the simulated unpolarized cross section}
\label{sec:D1}

In this section, we describe in more detail the Monte Carlo simulation of the unpolarized cross section and its fitting procedure, and we present the results of the parametrization of the unpolarized DiFF $D_1$. 

\subsection{The Monte Carlo simulation}
\label{sec:D1MC}
  
We used a {\tt PYTHIA} simulation~\cite{Sjostrand:2003wg} to study $(\pi^+ \pi^-)$ pairs with momentum fraction $z$ and invariant mass $M_h$ from $e^+ e^-$ annihilations at the Belle 
kinematics~\cite{courtesyBELLE}. The pair distribution should be described according to the unpolarized cross section of Eq.~\eqref{eq:dsig0-2} integrated in $\theta_2$ and $\theta$, since we assume the integration to be complete in the Monte Carlo sample. The actual expression of the cross section is 
\begin{equation}
\frac{d\sigma^0}{dz\,dM_h\,dQ^2} = 
\frac{4\pi \alpha^2}{Q^2}\,  \sum_q e_q^2\, D_1^q (z, M_h; Q^2) \; .
\label{eq:dsig0-MC}
\end{equation}

Events are generated with no cuts in acceptance. The data sample is based on a Monte Carlo integrated luminosity ${\cal L}_{\mathrm{MC}} = 647.26$ pb$^{-1}$ corresponding to $2.194\times 10^6$ events. The total number of produced pion pairs is $n_{\mathrm{tot}} = 1.040\times 10^6$, approximately one pair every two events. We use these numbers to normalize $D_1$, but the results for the Artru--Collins asymmetry (and, consequently, for $H_1^{\open} / D_1$) are independent of the normalization. 


The counts of pion pairs are collected in a bidimensional $40\times 50$ binning in $(z, M_h)$. The invariant mass is limited in the range $0.29 \leq M_h \leq 1.29$ GeV, the lower bound being given by the natural threshold $2 m_\pi$ and the upper cut excluding scarcely populated or frequently empty bins. Each pion pair is required to have a fractional energy $z \geq 0.2$ in order to focus only on pions coming from the fragmentation process. To avoid large mass corrections, we impose the condition 
\begin{equation}
\gamma_h \equiv \frac{2 M_h}{z Q} \ll 1 \; , 
\label{eq:zlim}
\end{equation}
which we in practice implement as $\gamma_h \leq 1/2$. 

For the fragmentation process $q\to (\pi^+ \pi^-) X$ in the range $0.29\leq M_h\leq 1.29$ GeV, the invariant mass distribution has a rich structure. The most prominent channels can be cast in two main categories,  three resonant channels and a ``continuum" (see the discussion around Fig.~2 in 
Ref.~\cite{Bacchetta:2006un}; see also 
Refs.~\cite{Acton:1992sa,Abreu:1992xx,Buskulic:1995gm,Fachini:2004jx}): 
\begin{itemize}
\item{} the production of $(\pi^+ \pi^-)$ pairs in relative $p$ wave via the decay of the $\rho$ resonance; it is the cleanest channel and is responsible for a peak in the invariant mass distribution at $M_h \sim 776$ MeV,

\item{} the production of $(\pi^+ \pi^-)$ pairs in relative $p$ wave via the decay of the $\omega$ resonance; it produces a sharp peak at $M_h \sim 783$ MeV but smaller than the previous one. However, the $\omega$ resonance has a large branching ratio for the decay into 
$(\pi^+ \pi^-) \pi^0$~\cite{Nakamura:2010zzi}. We include also this contribution after summing over the unobserved $\pi^0$; it generates a a broad peak roughly centered around $M_h \sim 500$ MeV,

\item{} the production of $(\pi^+ \pi^-)$ pairs via the decay of the $K_S^0$ resonance, which produces a very narrow peak at $M_h \sim 498$ MeV, 

\item{} everything else included in a channel which for convenience we call ``continuum" and we model as the fragmentation into an ``incoherent" pion pair.  
\end{itemize}
The fragmentation via the $\eta$ resonance also produces a peak overlapping with the $K_S^0$ one (plus a smaller hump at $M_h \sim 350$ MeV) but with less statistical weight. Hence, we will neglect this channel and we will neglect as well all other resonances which are not visible in the {\tt PYTHIA} 
output~\cite{Bacchetta:2006un}. 

In summary, the behaviour of the fragmentation into $(\pi^+ \pi^-)$ pairs with respect to their invariant mass will be simulated in four ways: three channels corresponding to the decay of the $\rho$, $\omega$, and $K^0_S$ resonances, and a channel that includes everything else (continuum). Using the Monte Carlo, we study each channel separately. For each channel, the flavor sum in Eq.~\eqref{eq:dsig0-MC} is decomposed in the contribution of $q = u,\, d,\, s,$ and $c$.


\subsection{Fitting the Monte Carlo simulation}
\label{sec:D1fit}

In the first step, for each channel $\mathrm{ch}= \mathrm{cont}, \, \rho, \, \omega, \, K$, and for each flavor $q = u,\, d,\, s,\, c$, we parametrize $D_{1,\mathrm{ch}}^q (z, M_h; Q_0^2)$ at the hadronic scale $Q_0^2=1$ GeV$^2$ taking inspiration from Refs.~\cite{Bacchetta:2006un,Bacchetta:2008wb,Bacchetta:2011ip}. For 
$(\pi^+ \pi^-)$ pairs, isospin symmetry and charge conjugation suggest that 
\begin{gather}
D_{1}^{u} = D_{1}^{d} = \overline{D}_1^u = \overline{D}_1^d \; ,  \label{eq:ass1}
\\
D_1^s = \overline{D}_1^s \; , \quad D_1^c = \overline{D}_1^c \; . \label{eq:ass2}
\end{gather}
The best fit of the Monte Carlo output at the Belle scale shows compatibility with both 
conditions~\eqref{eq:ass1} and~\eqref{eq:ass2} for all channels but for the $K^0_S \to (\pi^+ \pi^-)$ decay, where the choice $D_{1,K}^d \neq D_{1,K}^u$ is required. In general, we choose  $D_1^s$ to differ from  $D_1^u$ only in the $z$ dependence. 

The full analytic expression of $D_{1,\mathrm{ch}}^q (z, M_h; Q_0^2)$ can be found in 
appendix~\ref{sec:A}. Here, we illustrate the $z$ and $M_h$ dependence of $D_{1, \rho}^u$ as an example, since it displays enough general features that are common to most of the other channels. 
The function $D_{1, \rho}^u (z, M_h; Q_0^2)$ is described by 
\begin{equation}
\begin{split}
&D_{1, \rho}^u (z, M_h; Q_0^2) = (N_1^\rho)^2 z^{\alpha_1^\rho} (1-z)^{(\alpha_2^\rho)^2} 
(2 |\bm{R}| )^{(\beta_1^\rho)^2}  \\
&\quad \times \Bigg[ 
   \exp \left[ - P(\gamma_1^\rho, \gamma_2^\rho, \gamma_3^\rho, 0, -(\gamma_1^\rho+
                          \gamma_2^\rho+\gamma_3^\rho); z)\, M_h^2  \right]   \\
&\quad \qquad  
    \times \exp \left[ - P(\delta_1^\rho, 0, \delta_2^\rho, 0, 0; z) \right]  \\
&\qquad + (\eta_1^\rho)^2\, \mathrm{BW} (m_\rho, \Gamma_\rho; M_h) \Bigg]  \; , 
\label{eq:rho-u}
\end{split}
\end{equation}
where 
\begin{equation}
\begin{split}
P (a_1, a_2, a_3, a_4, a_5; x) &= a_1 \frac{1}{x} + a_2 + a_3 x + a_4 x^2 + a_5 x^3  \\
\mathrm{BW} (m, \Gamma; x) &= \frac{1}{(x^2-m^2)^2 + m^2 \Gamma^2}   \; .  
\label{eq:struct}
\end{split}
\end{equation}

The function BW is proportional to the modulus squared of a relativistic Breit--Wigner for the considered  resonant channel, and it depends on its mass and width. In this case of the $\rho \to (\pi^+ \pi^-)$ decay, it involves the fixed parameters $m_\rho = 0.776$ GeV and $\Gamma_\rho = 0.150$ GeV. The other ten parameters 
$(N_1^\rho,\, \alpha_1^\rho,\, \alpha_2^\rho,\, \beta_1^\rho,\, \gamma_1^\rho,\, \gamma_2^\rho,\, \gamma_3^\rho,\, \delta_1^\rho,\, \delta_2^\rho,\, \eta_1^\rho)$ are fitting parameters. In 
Eq.~\eqref{eq:rho-u}, the dependence on $z$ and $M_h$ is factorized, namely it can be represented as the product of two functions $f_1 (z)\, f_2(M_h)$, except for the exponential term 
$\exp [ P\, M_h^2]$, where $P$ is the polynomial depending only on $z$. A good fit of the Monte Carlo output can be reached only if the latter contribution is included. 

More generally, in every channel there is a factorized part where the $z$ dependence is of the kind $z^{\alpha_1} (1-z)^{\alpha_2}$ and the $M_h$ dependence is of the kind $2 |\bm{R}|^\beta$, with 
$|\bm{R}|$ given by Eq.~\eqref{eq:|R|}. The $\alpha_1, \, \alpha_2,$ and $\beta$, are fitting parameters. Then, the factorized part is multiplied by an unfactorizable contribution which can be generally represented as $\exp [ d_{\{\delta\}} (z) + h_{\{\lambda\}} (M_h) + f_{\{\gamma\}} (z M_h)]$. The functions $d, \, h, \, f,$ are typically polynomials depending also on sets of fitting parameters 
$\{\delta\}, \, \{\lambda\}, \, \{\gamma\},$ respectively. The appearance of the term 
$f_{\{\gamma\}} (z M_h)$ prevents the fitting function from assuming a factorized dependence in $z$ and $M_h$. The best fit of the Monte Carlo output requires a nonvanishing and important contribution from 
$f_{\{\gamma\}} (z M_h)$~\cite{Courtoy:2010qm}. For the resonant channels, the unfactorizable contribution is added to the modulus squared of a Breit--Wigner distribution in $M_h$ with the mass and width of the considered resonance and weighted with a fitting parameter $\eta$. The $K^0_S \to (\pi^+ \pi^-)$ decay requires a more elaborated analysis around the peak, since the resonance width is narrower than the width of the Monte Carlo bin (see appendix). 

The $\alpha_1,\, \alpha_2,\, \beta, \, \{\delta\}, \, \{\lambda\}, \, \{\gamma\}, \, \eta,$ sets of parameters (and the normalization $N$) can all depend on the selected channel and sometimes also on the flavor of the fragmenting quark. They are fixed by evolving each $D_{1,\mathrm{ch}}^q (z, M_h; Q_0^2)$ to the Belle scale $Q^2=100$ GeV$^2$ and then by fitting the Monte Carlo output for the unpolarized cross section 
$d\sigma^0$ of Eq.~\eqref{eq:dsig0-MC} for each channel $\mathrm{ch}$ at $Q^2=100$ GeV$^2$ by minimizing
\begin{equation}
\chi^2_{\mathrm{ch}} = \sum_q \sum_{ij} 
               \frac{\left( N_{ij}^{\mathrm{ch},\, q} - 
                        {\cal L}_{\mathrm{MC}} \, ( d\sigma^{0\, q}_{\mathrm{ch}} )_{ij} \right)^2}
                       {{\cal L}_{\mathrm{MC}} \, ( d\sigma^{0\, q}_{\mathrm{ch}} )_{ij} } \; , 
\label{eq:chi2}
\end{equation}
where $N_{ij}^{\mathrm{ch},\, q}$ is the number of pion pairs produced in the simulation by the flavor $q$ in the channel $\mathrm{ch}$ in the bin $(z_i, \, M_{h\, j})$. The $( d\sigma^{0\, q}_{\mathrm{ch}} )_{ij}$ is the fitting unpolarized cross section for the specific flavor $q$ and channel $\mathrm{ch}$, integrated over the bin $(z_i, \, M_{h\, j})$ of width $(\Delta z, \, \Delta M_h)$, i.e. 
\begin{equation}
\begin{split}
( d\sigma^{0\, q}_{\mathrm{ch}} )_{ij} &\equiv \int_{z_i}^{z_i+\Delta z} dz 
\int_{M_{h\, j}}^{M_{h\, j}+\Delta M_h} dM_h \, \frac{d\sigma^{0\, q}_{\mathrm{ch}}}{dz\,dM_h\,dQ^2} 
\\
&= \frac{4\pi \alpha^2}{Q^2}\,  e_q^2 \\
&\times \int_{z_i}^{z_i+\Delta z} dz \int_{M_{h\, j}}^{M_{h\, j}+\Delta M_h} dM_h \, 
D_{1,\mathrm{ch}}^q (z, M_h; Q^2) \; .
\label{eq:dsig0-MC-bin}
\end{split}
\end{equation}
In order to make the computation less heavy, we have approximated the integral in the above equation with the $d\sigma^{0\, q}_{\mathrm{ch}}$ evaluated in the central value of the bin $(z_i,\, M_{h\, j})$, and multiplied by $\Delta z \Delta M_h$. We have checked that this approximation introduces negligible systematic errors. Evolution effects are  calculated using the {\tt HOPPET} code~\cite{Salam:2008qg}. Splitting functions have been considered at LO. Gluons are generated only radiatively, because a nonvanishing gluon DiFF $D_1^g$ at the starting scale $Q_0^2$ would be largely unconstrained. Nevertheless, we reach good fits for all channels (see Tab.~\ref{tab:chi2}). 


\begin{table}[h]
\begin{tabular}{|c||c|c|c|c|c|}
\hline
 & cont & $\rho$  & $\omega$ &  $K^0_S$ & global  \\
\hline
\hline
$\chi^2$/dof & 1.69 & 1.28 & 1.68 & 1.85  &  1.62  \\
\hline
\end{tabular}
\caption{The $\chi^2$/dof obtained by fitting the simulated yield of $(\pi^+\pi^-)$ pairs produced either directly (continuum), or via the $\rho, \, \omega,$ or $K^0_S$ resonances, and the global one.} 
\label{tab:chi2} 
\end{table} 

The $\chi^2_{\mathrm{ch}}$ minimization is performed using {\tt MINUIT}, separately for each channel, on a grid of $40\times 50 \times 4$ bins in $(z_i, \, M_{h\, j}, \, \mathrm{flavor})$ (the actual dimension of the grid is slightly smaller because of the constraint in Eq.~\eqref{eq:zlim}). In Tab.~\ref{tab:chi2}, we list the values of the $\chi^2_{\mathrm{ch}}$ per degree of freedom ($\chi^2_{\mathrm{ch}} /$ dof) for each channel as well as of the global one, obtained from their average weighted over the fraction of total degrees of freedom. The continuum can be represented with 17 parameters. Each of the $\rho$ and 
$\omega$ channels involves 20 parameters, while the $K^0_S$ resonance 22 ones. Their best values are listed in the appendix, together with their statistical errors. As an example, in Tab.~\ref{tab:rho-ud} we list the best values of the fitting parameters in Eq.~\eqref{eq:rho-u} together with their statistical errors, corresponding to $\Delta \chi^2 = 1$. The theoretical uncertainty on $D_{1,\mathrm{ch}}^q$ at $Q_0^2$ and on $d\sigma^0$ at the Belle scale are calculated using the covariant error matrix from {\tt MINUIT} and the standard formula for error propagation.

\begin{table}[h]
\begin{tabular}{|c||c|c|}
\hline
$\rho$ &  &   \\
\hline
\hline
$u=d$ & $N_1^\rho  =0.209\pm 0.011$ & $\beta_1^\rho = 0.999\pm 0.013$ \\
\hline
& $\alpha_1^\rho = 0.104\pm 0.025$ & $\alpha_2^\rho = -1.2095\pm 0.0078$  \\
\hline
 & $\gamma_1^\rho = 4.045\pm 0.173$ & $\gamma_2^\rho = -15.679\pm 0.870$ \\
 \hline
 & $\gamma_3^\rho = 20.582\pm 1.205$ &  $\eta_1^\rho =  1.103\pm 0.057$ \\
\hline
 & $\delta_1^\rho = -1.067 \pm 0.023$  &  $\delta_2^\rho = -1.357 \pm 0.140$  \\ 
\hline
\end{tabular}
\caption{Best-fit parameters for $D_{1, \rho}^u (z, M_h; Q_0^2)$ from Eq.~\eqref{eq:rho-u}. The errors correspond to $\Delta \chi^2 = 1$.} 
\label{tab:rho-ud} 
\end{table} 


\subsection{Results for $D_1$}
\label{sec:D1results}

\begin{figure}[h]
\begin{center}
\includegraphics[width=8.5cm]{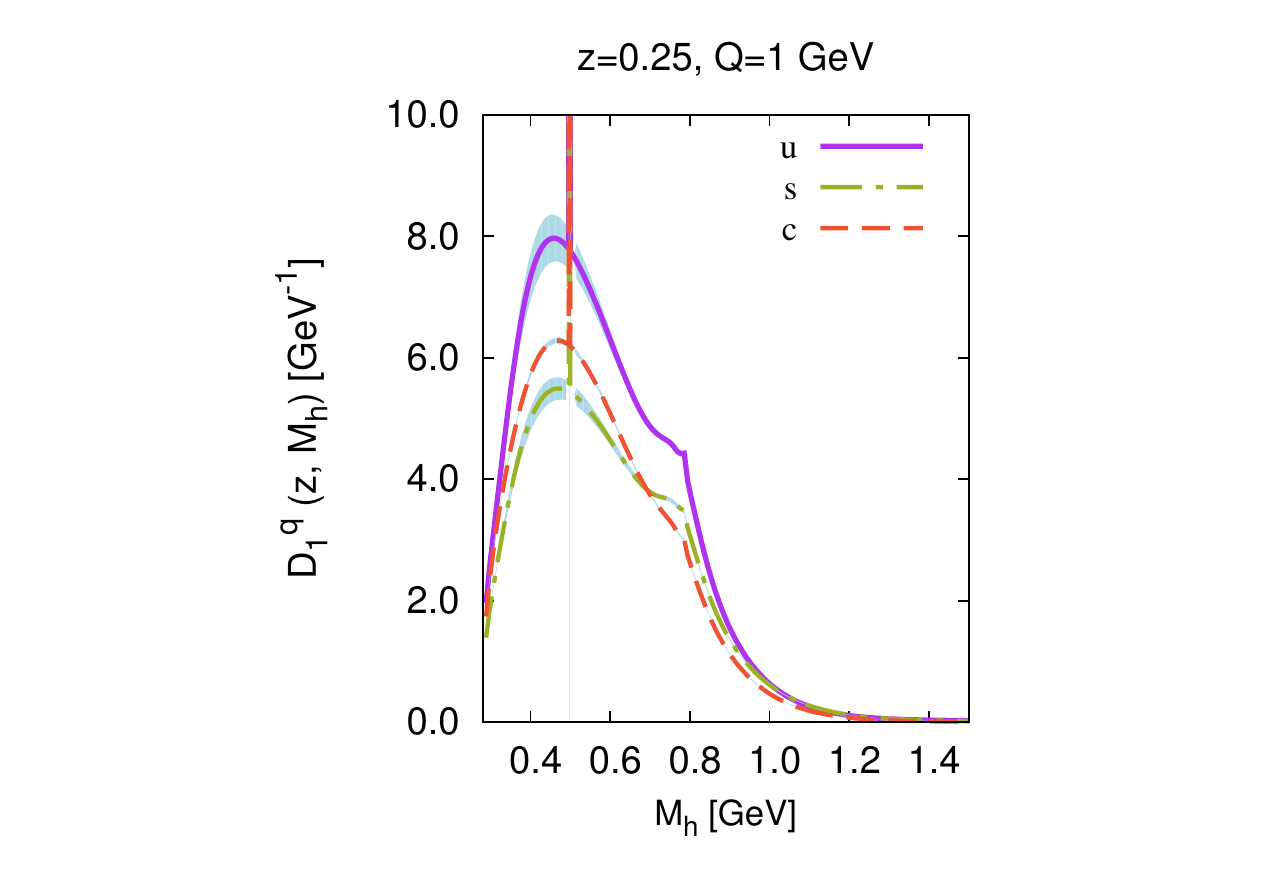}\\
\includegraphics[width=8.5cm]{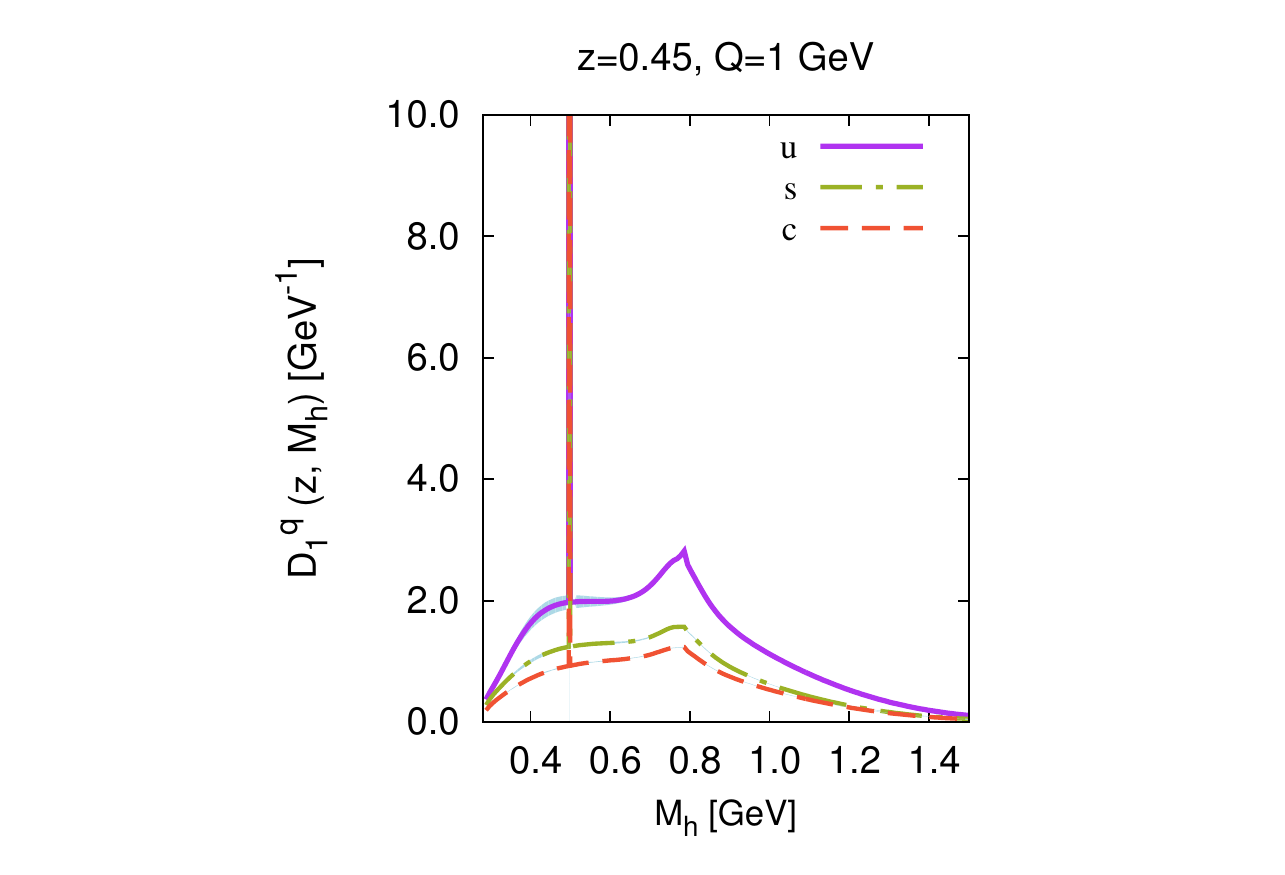}\\
\includegraphics[width=8.5cm]{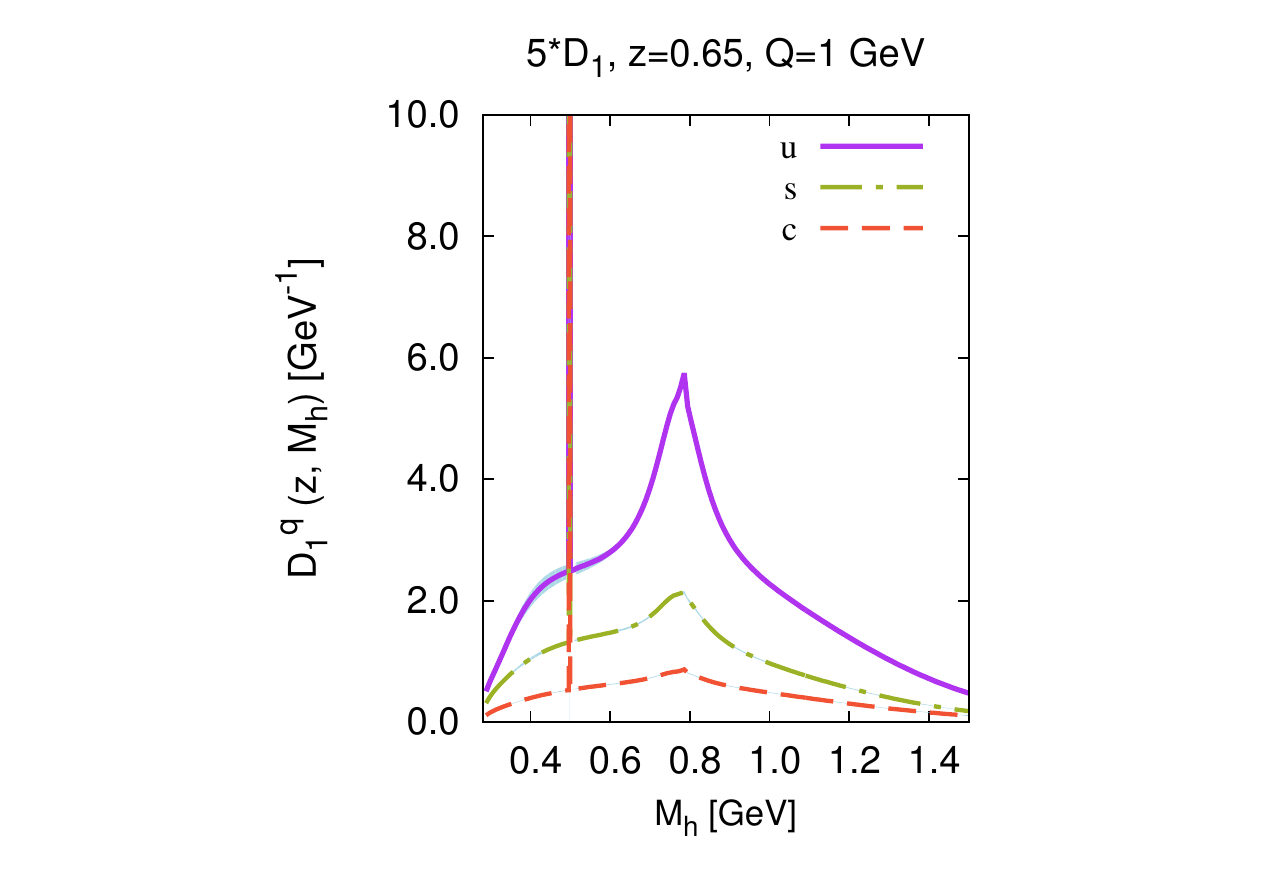}
\end{center}
\caption{The $D_1^q(z, M_h; Q_0^2)$, summed over all channels, as a function of $M_h$ for $z=0.25, \, 0.45,$ and $0.65$ (from top to bottom) at the hadronic scale $Q_0^2=1$ GeV$^2$. Solid, dot-dashed, and dashed, curves correspond to the contribution of the flavors $u=d, \, s,$ and $c$, respectively.}
\label{fig:D1Mh}
\end{figure}

\begin{figure}[h]
\begin{center}
\includegraphics[width=8.5cm]{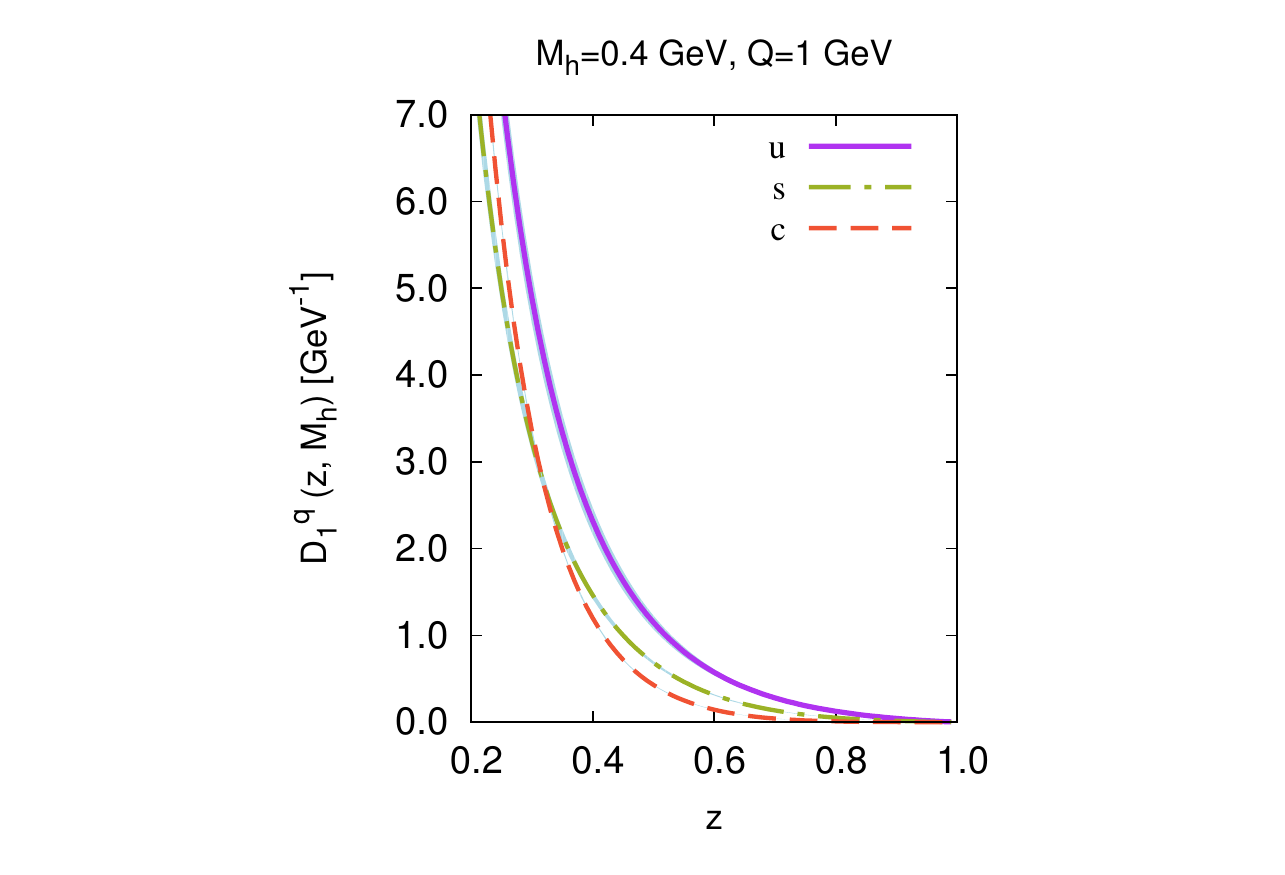}\\
\includegraphics[width=8.5cm]{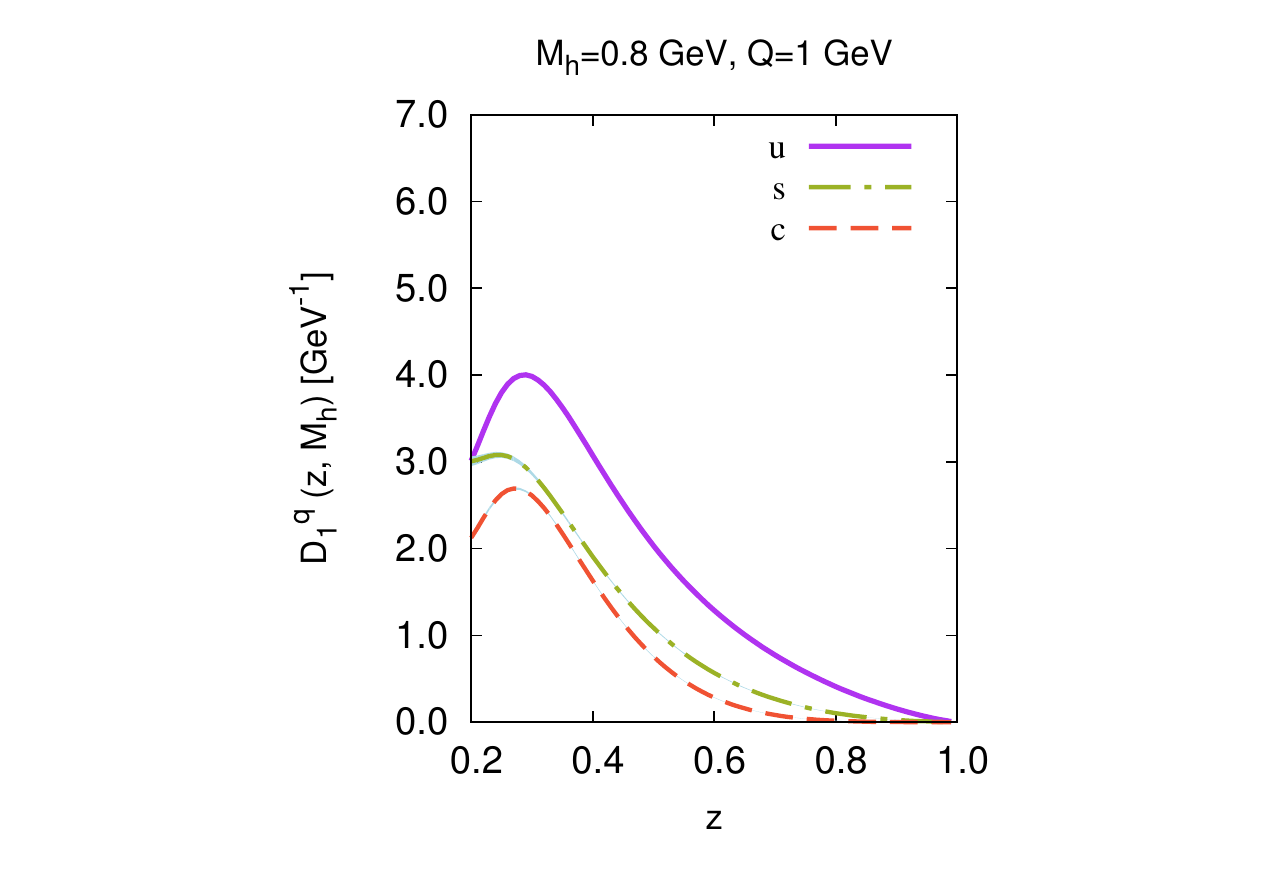}\\
\includegraphics[width=8.5cm]{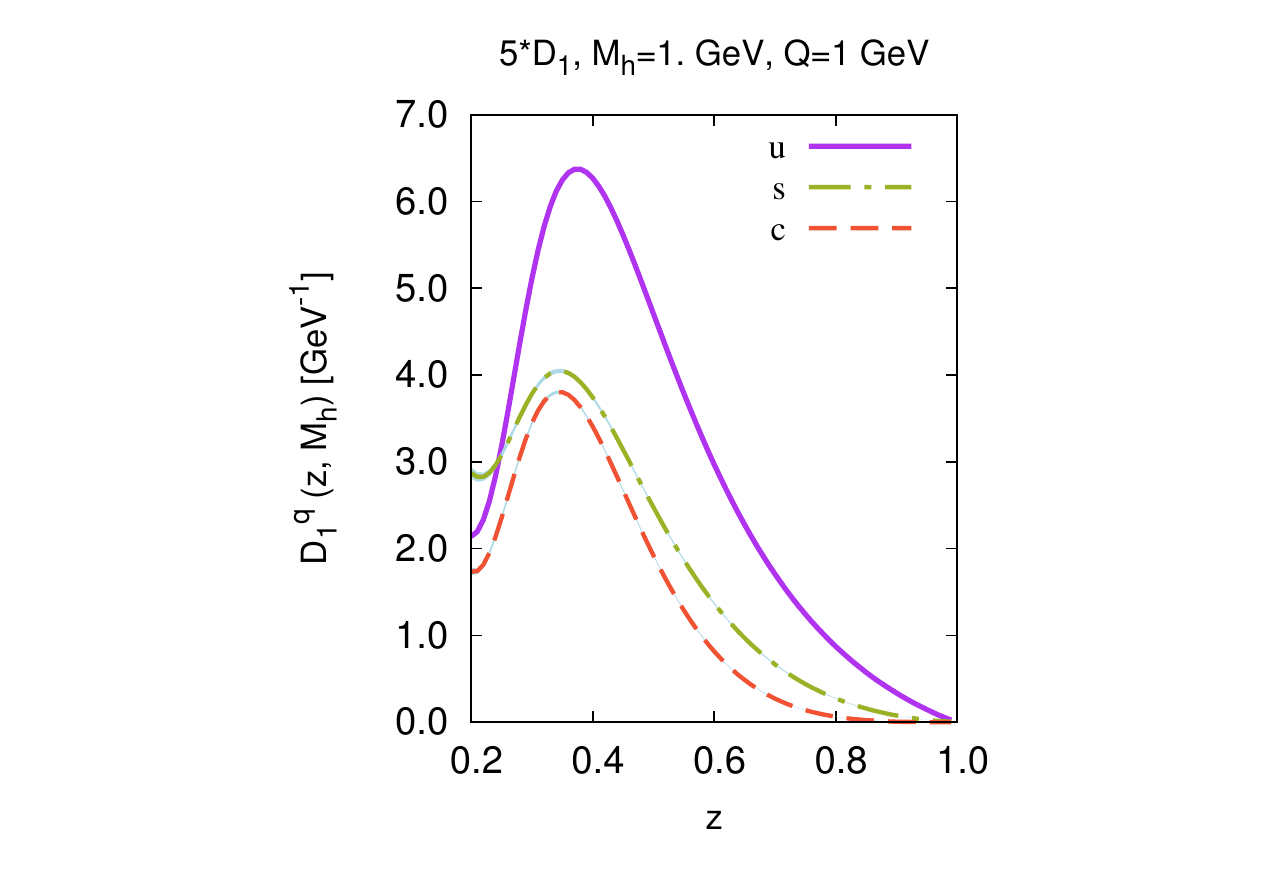}
\end{center}
\caption{The $D_1^q(z, M_h; Q_0^2)$, summed over all channels, as a function of $z$ for $M_h=0.4, \, 0.8,$ and $1$ GeV (from top to bottom) at the hadronic scale $Q_0^2=1$ 
GeV$^2$. Same notations as in previous figure.}
\label{fig:D1z}
\end{figure}

In Fig.~\ref{fig:D1Mh}, we show $D_1^q(z, M_h; Q_0^2)$, summed over all channels, as a function of $M_h$ for $z=0.25, \, 0.45,$ and $0.65$ (from top to bottom) at the starting scale $Q_0^2=1$ 
GeV$^2$. For each panel, the solid, dot-dashed, and dashed, curves correspond to the contribution of the flavors $u, \, s,$ and $c$, respectively. The $d$ contribution is identical to the $u$ one, according to Eq.~\eqref{eq:ass1}, but for the $K^0_S \to \pi^+ \pi^-$ channel, where the difference is anyway small. We recall that at this scale we assume no contribution from the gluon. The DiFFs are normalized using the Monte Carlo luminosity ${\cal L}_{\mathrm{MC}}$, although the overall normalization will not influence the results of the next sections. In the top panel, we can distinguish the narrow peak due to the $K^0_S$ resonance on top of a large hump, due to the superposition of the contributions coming from the continuum and from the 
$\omega \to (\pi^+ \pi^-) \pi^0$ decay. At $M_h=0.77$ GeV, we clearly see the peak of the $\rho$ resonance. Instead, the peak of the $\omega \to (\pi^+ \pi^-)$ decay is hardly visible. Moving from top to bottom, we can appreciate how the relative importance of the $\rho$ channel increases over the other ones as $z$ increases. 

In Fig.~\ref{fig:D1z}, we show $D_1^q(z, M_h; Q_0^2)$, summed over all channels, as a function of $z$ for $M_h=0.4, \, 0.8,$ and $1$ GeV (from top to bottom) at the starting scale $Q_0^2=1$ 
GeV$^2$. Notations are the same as in the previous figure. It is worth noting the relatively high importance of the charm contribution, especially at low $z$ for low and intermediate values of 
$M_h$. 

\begin{figure}[ht]
\begin{center}
\includegraphics[width=8.2cm]{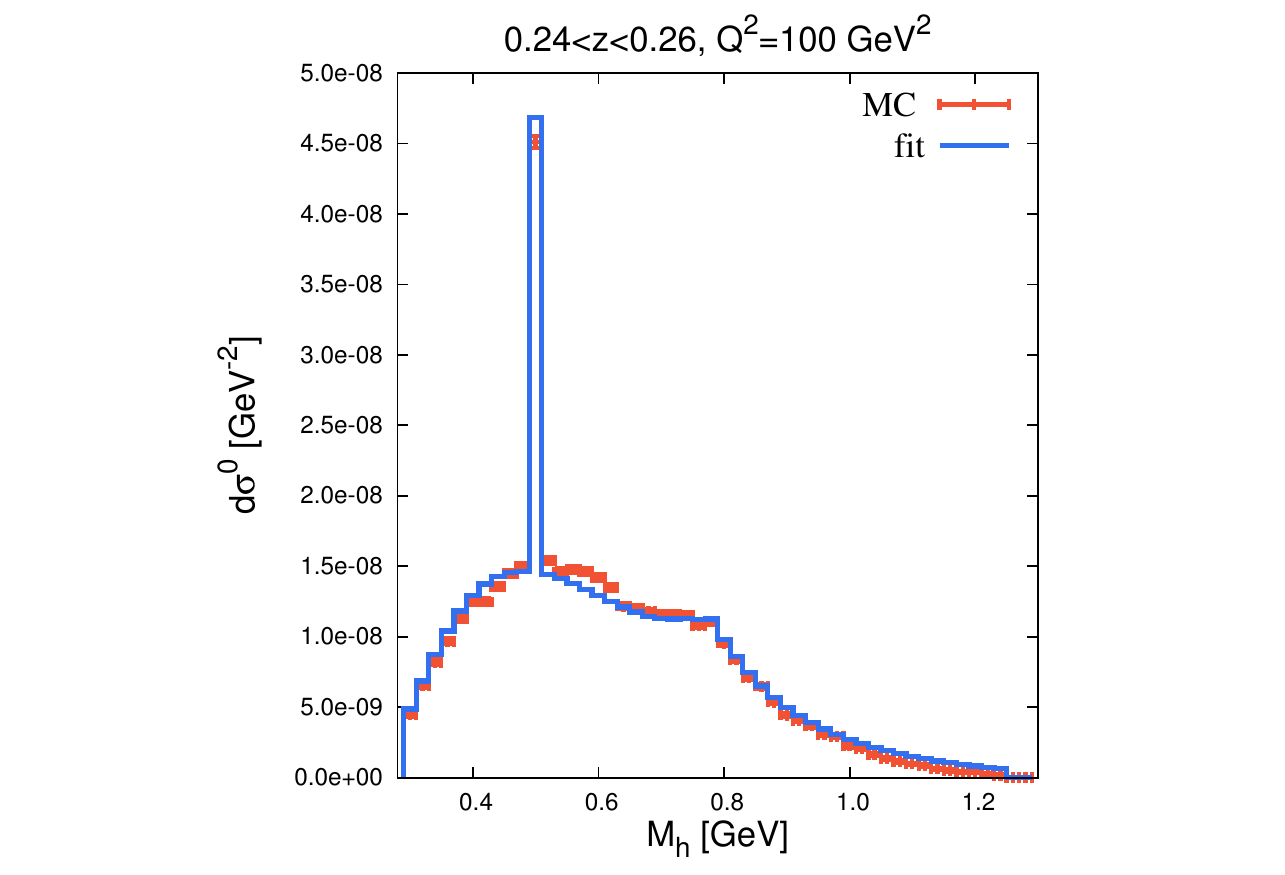}\\
\includegraphics[width=8.2cm]{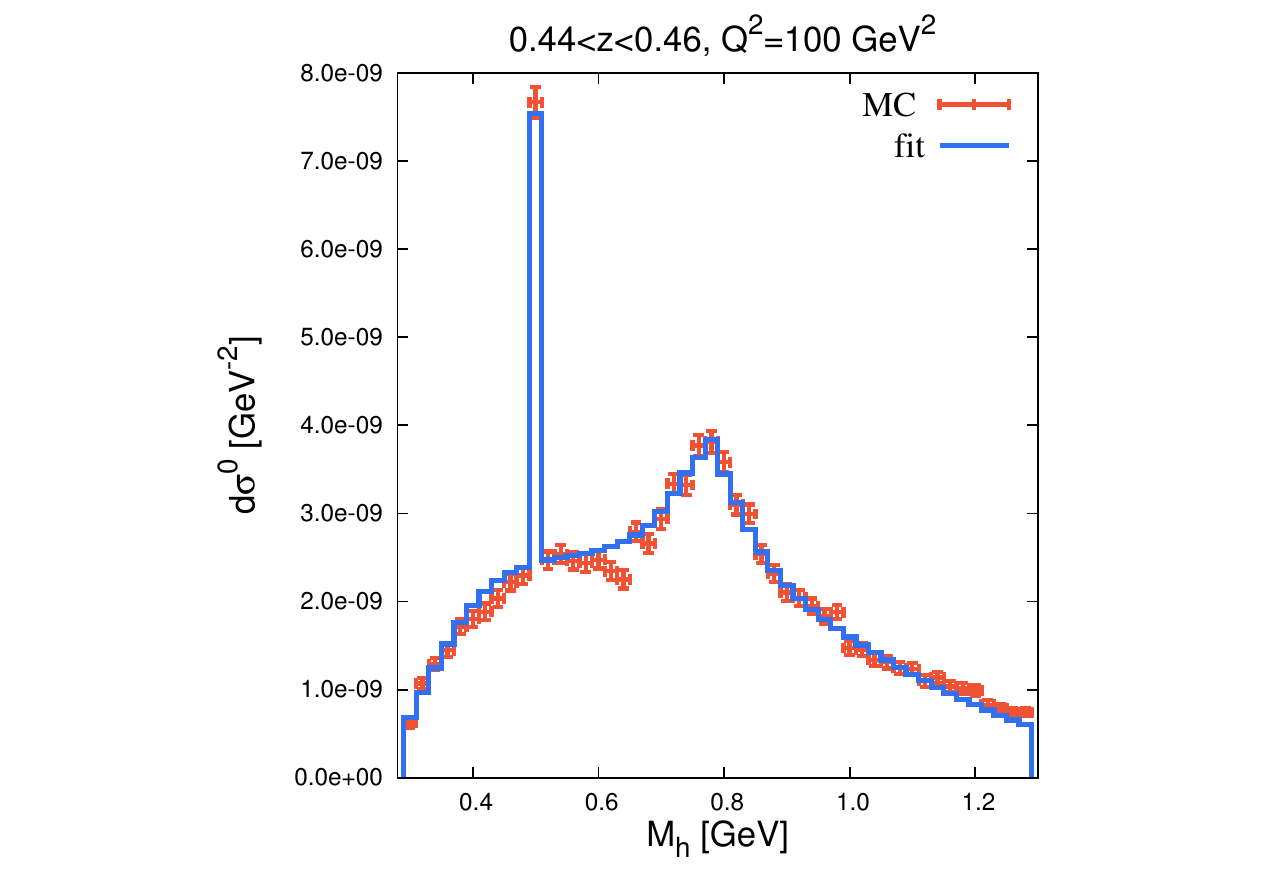}\\
\includegraphics[width=8.2cm]{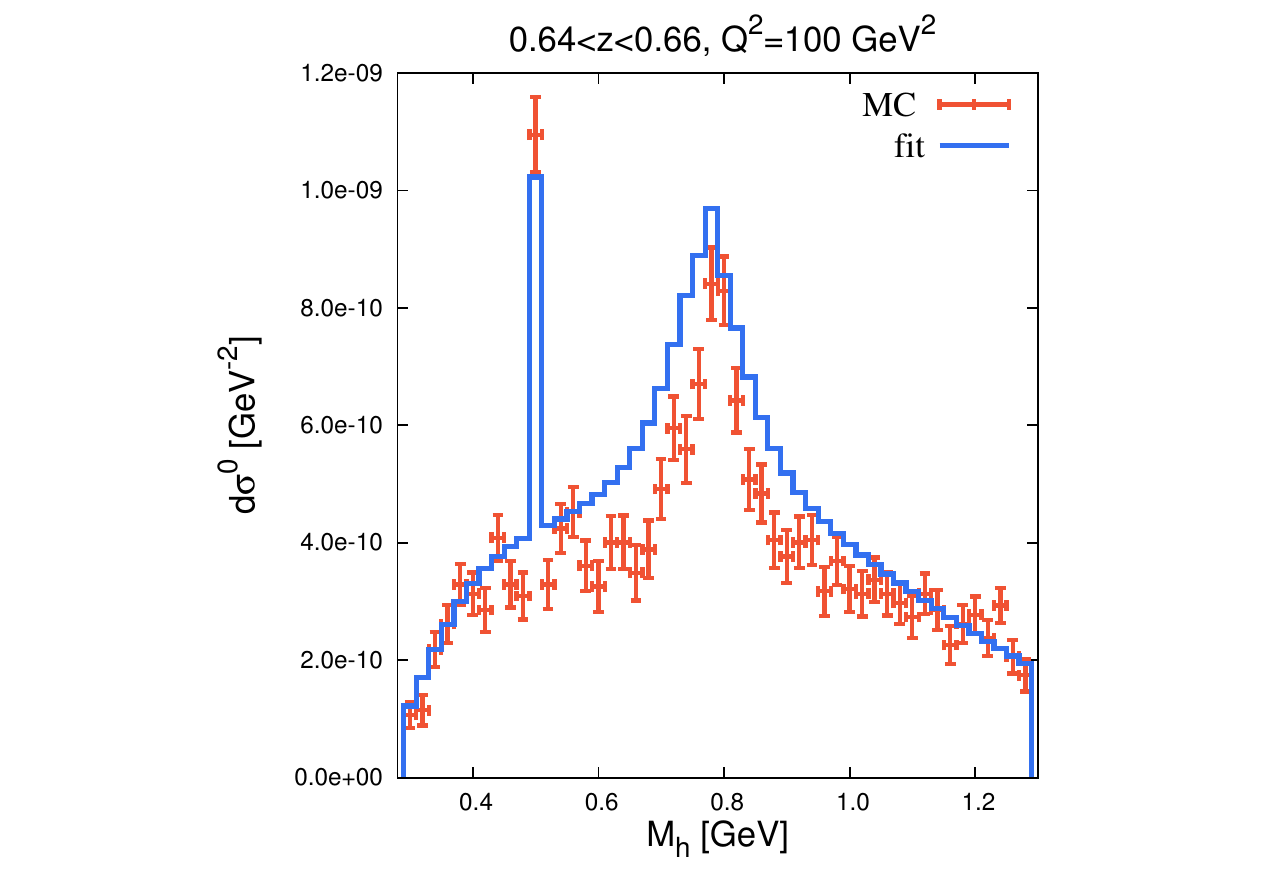}
\end{center}
\vspace{-0.5cm}
\caption{The unpolarized cross section $d\sigma^0$ at $Q^2=100 $ GeV$^2$ as a function of $M_h$ for the three bins $0.24\leq z\leq 0.26, \, 0.44\leq z\leq 0.46, \, 0.64\leq z\leq 0.66$ (from top to bottom). Histograms for the fitting formula of Eq.~\eqref{eq:dsig0-MC-bin}, summed over all flavors and channels, and integrated in each $M_h$ bin. Points with error bars for the simulated observable with statistical errors. The figure serves  only for illustration purposes. For the description of the actual fitting procedure, see details in the text, particularly around Eqs.~\eqref{eq:chi2} and~\eqref{eq:dsig0-MC-bin}.}
\label{fig:sig0Mh}
\end{figure}

\begin{figure}[ht]
\begin{center}
\includegraphics[width=8.2cm]{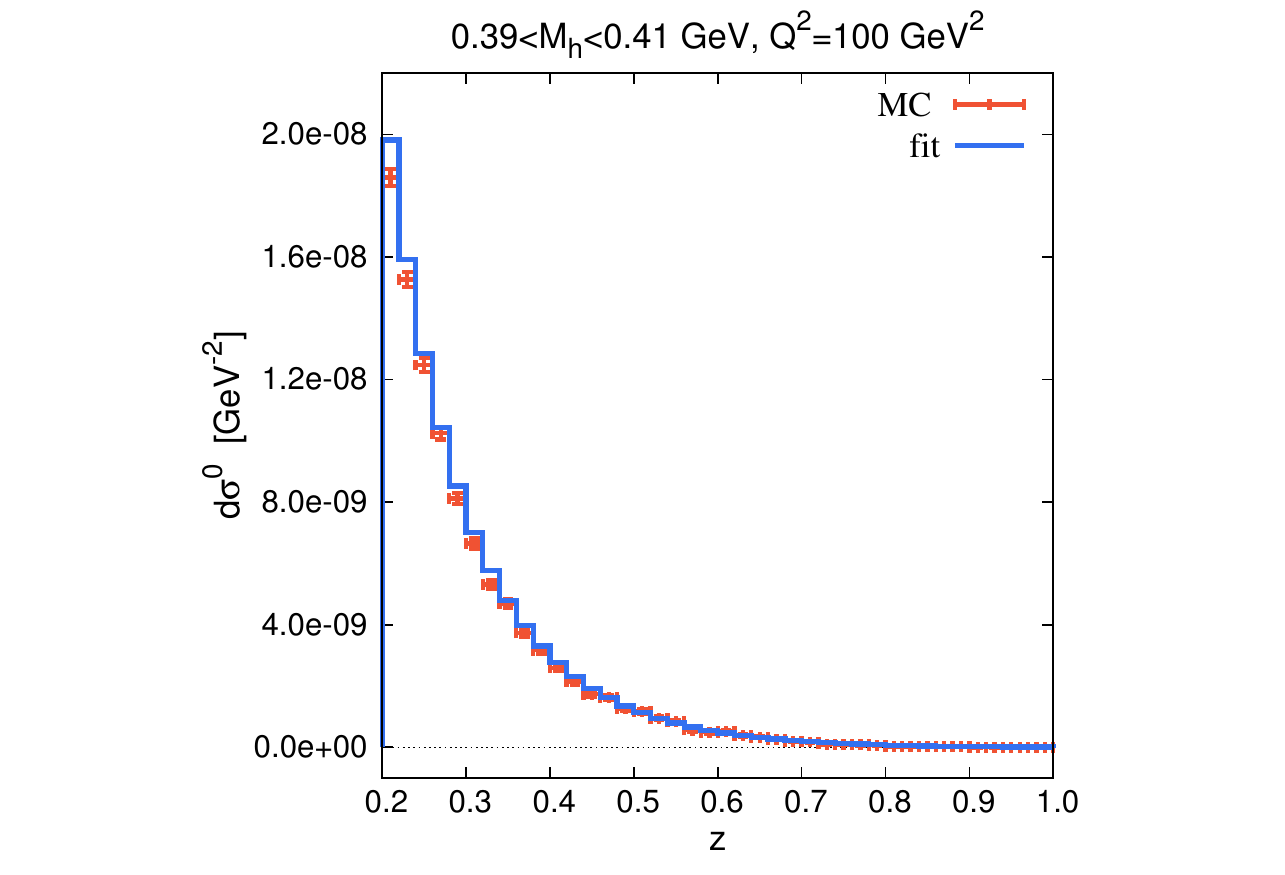}\\
\includegraphics[width=8.2cm]{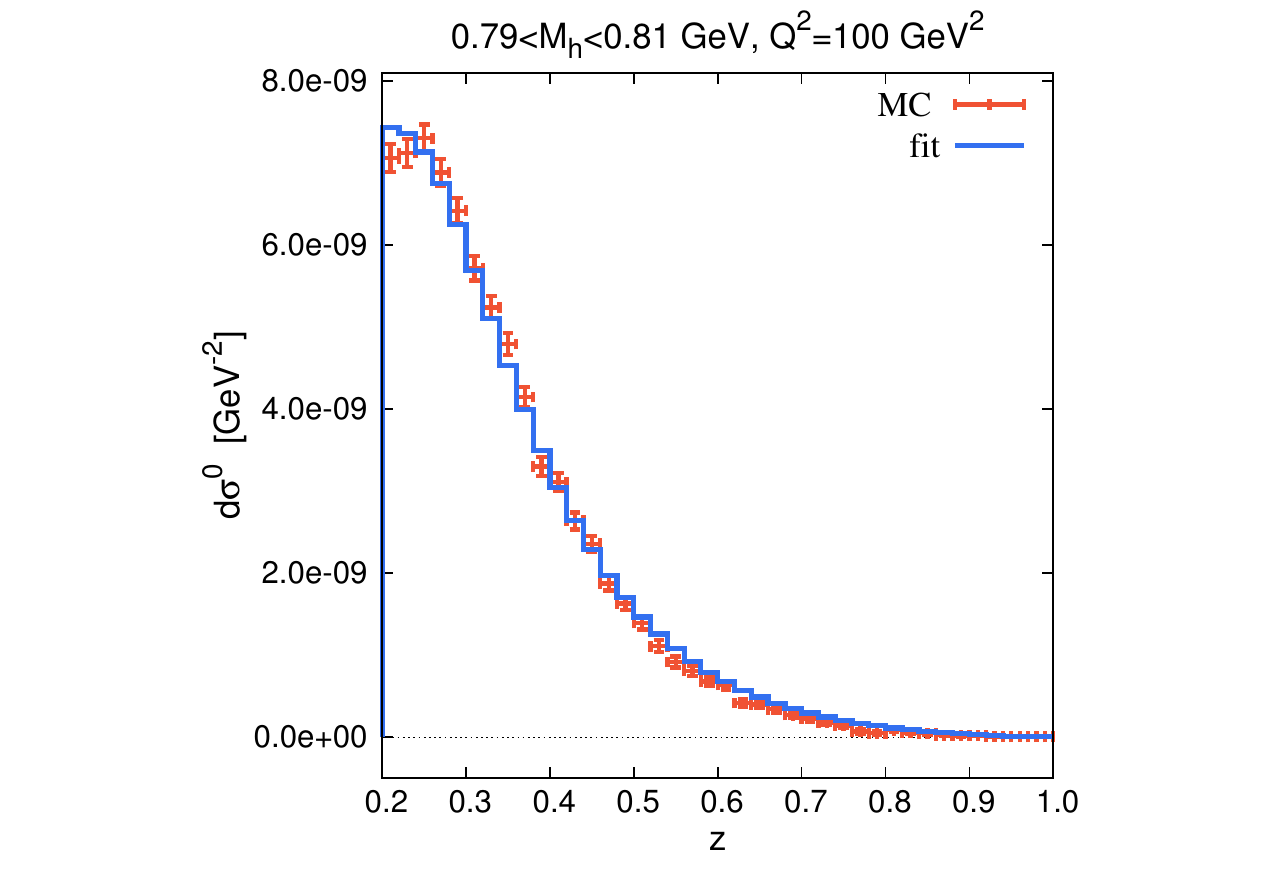}\\
\includegraphics[width=8.2cm]{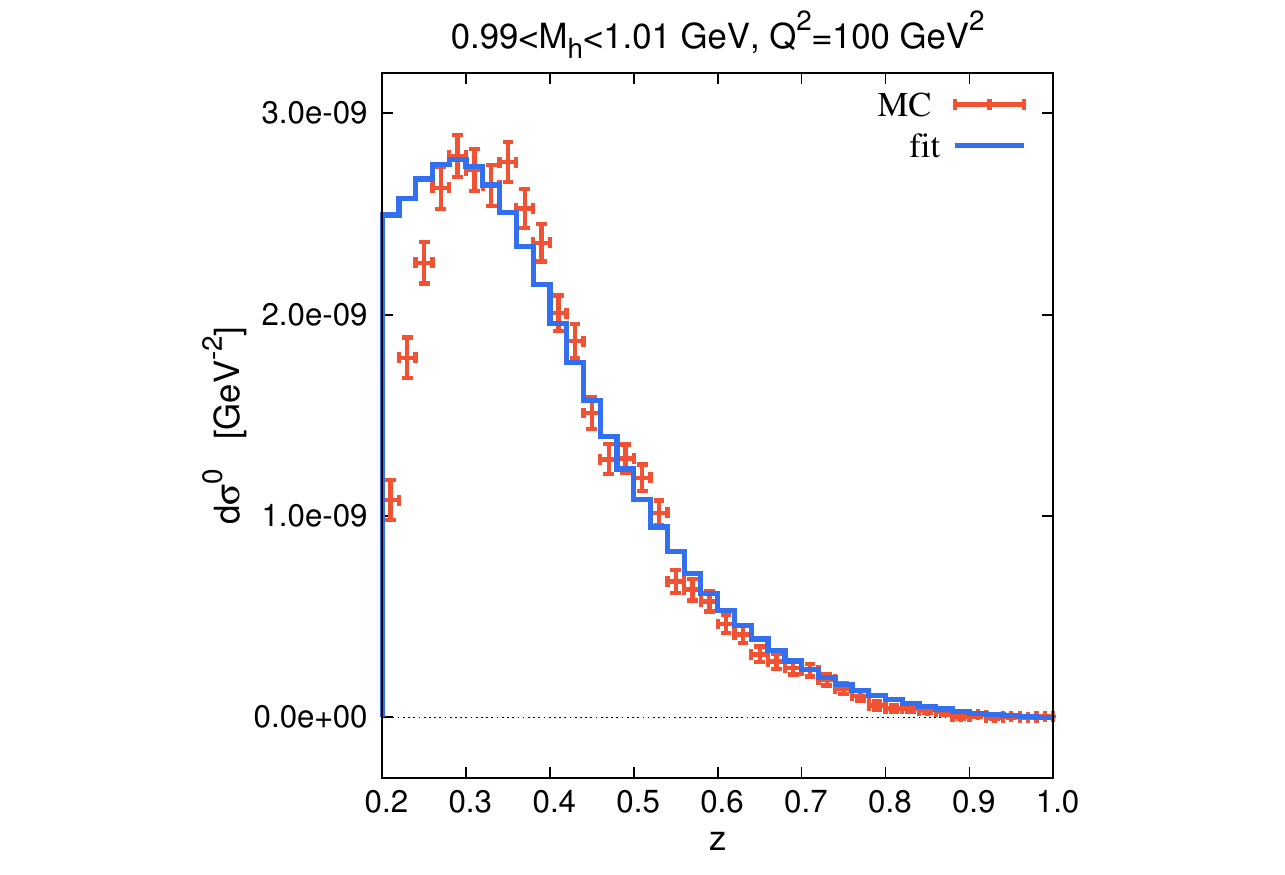}
\end{center}
\caption{The unpolarized cross section $d\sigma^0$ at $Q^2=100$ GeV$^2$ as a function of $z$ for the three bins $0.39\leq M_h\leq 0.41,\, 0.79\leq M_h\leq 0.81, \, 0.99\leq M_h\leq 1.01$ GeV (from top to bottom). Same notations as in the previous figure. The figure serves  only for illustration purposes. For the description of the actual fitting procedure, see details in the text, particularly around Eqs.~\eqref{eq:chi2} 
and~\eqref{eq:dsig0-MC-bin}.}
\label{fig:sig0z}
\end{figure}

In Fig.~\ref{fig:sig0Mh}, the points with error bars are the numbers $N_{ij}$ of pion pairs produced by the simulation in the bin $(z_i, \, M_{h\, j})$, summed over all flavors and channels and divided by the Monte Carlo luminosity ${\cal L}_{\mathrm{MC}}$; i.e., they represent the simulated experimental unpolarized cross section with errors defined in Eq.~\eqref{eq:chi2}. The histograms refer to 
$( d\sigma^{0\, q}_{\mathrm{ch}} )_{ij}$ in Eq.~\eqref{eq:dsig0-MC-bin} summed over all flavors and channels, i.e., to the fitting unpolarized cross section evolved at the Belle scale $Q^2 = 100$ 
GeV$^2$. In reality, we have independently fitted each of the four channels. For illustration purposes, here we show the plots in the $M_h$ bins only for the three bins 
$0.24\leq z\leq 0.26, \, 0.44\leq z\leq 0.46, \, 0.74\leq z\leq 0.76$ (from top to bottom, respectively) after summing upon all flavors and channels. The agreement between the histogram of theoretical predictions and the points for the simulated experiment confirms the good quality of the fit. As in Fig.~\ref{fig:D1Mh}, going from top to bottom panels one can appreciate the modifications with changing 
$z$ of the relative weight among the various channels active in the invariant mass distribution (kaon peak, 
$\rho$ peak, broad continuum, etc..). 

In Fig.~\ref{fig:sig0z}, the fitting $( d\sigma^{0\, q}_{\mathrm{ch}} )_{ij}$ and simulated 
$N_{ij} / {\cal L}_{\mathrm{MC}}$ unpolarized cross sections, summed over all flavors and channels, are now plotted as functions of the $z$ bins for the three bins 
$0.39\leq M_h\leq 0.41,\, 0.79\leq M_h\leq 0.81, \, 0.99\leq M_h\leq 1.01$ GeV (from top to bottom) in the same conditions and with the same notations as in the previous figure. The agreement remains very good but for few bins at low $z$ at the highest considered $M_h$, and confirms the quality of the extracted parametrization of the unpolarized DiFF.


\section{Extraction of $H_1^{\open}$ from measured Artru--Collins asymmetry}
\label{sec:H1}

We now consider the Artru--Collins asymmetry of Eq.~\eqref{eq:ACasy}. Since we cannot integrate away the $\theta_2,\, \theta,$ and $\overline{\theta}$ angles in the experimental acceptance, we will consider their average values in each experimental bin. As such, Eq.~\eqref{eq:ACasy} corresponds to the experimental $a_{12R}$ in Ref.~\cite{Vossen:2011fk}. 

It is convenient to define also the following quantities
\begin{equation}
\begin{split}
n_q(Q^2) &= \int_{0.2}^1 dz \int_{2 m_\pi}^2 dM_h \, D_1^q (z, M_h; Q^2)   \\
n_q^\uparrow (Q^2) &= \int_{0.2}^1 dz \int_{2 m_\pi}^2 dM_h \, \frac{|\bm{R}|}{M_h}\, H_{1,sp}^{\open\, q}(z,M_h; Q^2) \; .
\label{eq:nDiFF}
\end{split}
\end{equation}
Then, the Artru--Collins asymmetry can be simplified to
\begin{equation}
\begin{split} 
A(z, M_h; Q^2) &=  - \frac{\langle \sin^2 \theta_2 \rangle}{\langle 1+\cos^2 \theta_2\rangle} \,
\langle \sin\theta \rangle \langle \sin\overline{\theta}\rangle \\
&\times \frac{|\bm{R}|}{M_h} \, \frac{\sum_q e_q^2 \, H_{1, sp}^{\open q}(z, M_h; Q^2)\,
                                                                    n_q^\uparrow (Q^2)}
                                                              {\sum_q e_q^2\, D_1^q (z, M_h; Q^2) \, n_q(Q^2)}  \\
&\equiv - \frac{\langle \sin^2 \theta_2 \rangle}{\langle 1+\cos^2 \theta_2\rangle} \,
\langle \sin\theta \rangle \langle \sin\overline{\theta}\rangle \\
&\times \frac{|\bm{R}|}{M_h} \, \frac{\sum_q e_q^2 \, H_{1, sp}^{\open q}(z, M_h; Q^2)\,
                                                                    n_q^\uparrow (Q^2)}
                                                              {D(z, M_h; Q^2)} \; ,
\label{eq:ACasyMC}
\end{split} 
\end{equation} 
where we understand that $\overline{n}_q(Q^2) = n_q(Q^2)$ (due to Eqs.~\eqref{eq:ass1}, 
\eqref{eq:ass2}), $\overline{n}_q^\uparrow (Q^2) = - n_q^\uparrow (Q^2)$ (see the following 
Eqs.~\eqref{eq:ass3}, \eqref{eq:ass4}), and we have defined 
\begin{equation}
\begin{split}
&D(z, M_h; Q^2) = \\
&\quad  \frac{4}{9}\, D_1^u (z, M_h; Q^2) \, n_u (Q^2) + 
\frac{1}{9}\, D_1^d (z, M_h; Q^2) \, n_d (Q^2) \\
&\quad + \frac{1}{9}\, D_1^s (z, M_h; Q^2) \, n_s (Q^2) + 
\frac{4}{9}\, D_1^c (z, M_h; Q^2) \, n_c (Q^2) \; .
\label{eq:ACasyden}
\end{split}
\end{equation}

Isospin symmetry and charge conjugation can be applied also to the polarized fragmentation into 
$(\pi^+ \pi^-)$ pairs such that~\cite{Bacchetta:2006un,Bacchetta:2008wb,Bacchetta:2011ip} 
\begin{gather}
H_{1}^{\open,\, u} = -H_{1}^{\open,\, d} = -\overline{H}_1^{\open,\, u} = \overline{H}_1^{\open,\, d}  \; ,  \label{eq:ass3}
\\
H_1^{\open,\, s} = -\overline{H}_1^{\open,\, s}  = H_1^{\open,\, c} = -\overline{H}_1^{\open,\, c} = 0 \; . \label{eq:ass4}
\end{gather}
These relations should hold for all channels but for the $K^0_S$ resonance. However, pion pairs produced in the $K^0_S$ decay are in the relative $s$ wave, and with our assumptions there are no $p$ wave contributions to interfere with. Therefore, we assume $H_{1,sp}^{\open,\, q} \approx 0$ for the $K^0_S$ channel, such that Eqs.~\eqref{eq:ass3} and~\eqref{eq:ass4} are valid in general throughout our analysis. 

Using these symmetry relations, we can further manipulate Eq.~\eqref{eq:ACasyMC} and define 
\begin{equation}
\begin{split} 
&H(z, M_h; Q^2) = - \frac{\langle 1+\cos^2 \theta_2\rangle}{\langle \sin^2 \theta_2 \rangle}\,  
     \frac{9}{5} \,\frac{1}{\langle \sin\theta \rangle\,\langle \sin\overline{\theta}\rangle} \\
&\mbox{\hspace{1cm}} \times D(z, M_h; Q^2) \, A(z, M_h; Q^2) \\
&\quad \equiv
     \frac{|\bm{R}|}{M_h} \, H_{1, sp}^{\open u}(z, M_h; Q^2)\, n_u^\uparrow (Q^2) \; ,
\label{eq:ACasyfit}
\end{split} 
\end{equation} 
where 
\begin{equation}
\int_{0.2}^1 dz \int_{2m_\pi}^2 dM_h \, H(z, M_h, Q^2) = [ n_u^\uparrow (Q^2)]^2 \; . 
\label{eq:Hnorm}
\end{equation}

Our strategy is the following. At the hadronic scale $Q_0^2=1$ GeV$^2$, we parametrize 
$H(z, M_h; Q_0^2)$. Then, we evolve it using the {\tt HOPPET} code~\cite{Salam:2008qg}, suitably extended to include LO chiral-odd splitting functions. At the Belle scale of $Q^2=100$ GeV$^2$, we fit the function $H$ using Eq.~\eqref{eq:ACasyfit}, i.e. employing bin by bin the measured 
Artru--Collins asymmetry $A$, the average values of angles 
$\theta_2, \, \theta, \, \overline{\theta}$, and the asymmetry denominator $D$. The latter is obtained from Eqs.~\eqref{eq:ACasyden} and~\eqref{eq:dsig0-MC-bin} by fitting the Monte Carlo simulation of the unpolarized cross section. The final step consists in the identification 
\begin{equation}
\begin{split}
\frac{|\bm{R}|}{M_h} \, &H_{1, sp}^{\open u}(z, M_h; Q^2) = \\
 &\frac{H(z, M_h, Q^2)}{\left( \int_{0.2}^1 dz \int_{2m_\pi}^2 dM_h \, H(z, M_h, Q^2) \right)^{1/2}} \; .
\label{eq:H1}
\end{split}
\end{equation}
This result is possible because of the symmetry relations~\eqref{eq:ass3} and~\eqref{eq:ass4}. In fact, the chiral-odd splitting functions do not mix quarks with gluons in the evolution, but they can mix quarks with different flavors. However, Eqs.~\eqref{eq:ass3} and~\eqref{eq:ass4} imply that only the flavors $u$ or $d$ are actually active in the asymmetry and they are the same. Consequently, the factorized expression of $H$ in Eq.~\eqref{eq:ACasyfit} is preserved with changing $Q^2$, thus justifying Eq.~\eqref{eq:H1}.



\subsection{Fitting the experimental data}
\label{sec:H1plan}

The experimental data on the Artru--Collins asymmetry are organized in three different grids: a 
$9\times 9$ one in $(z, \overline{z})$, a $8\times 8$ one in $(M_h, \overline{M}_h)$, and a 
$8\times 8$ one in $(z, M_h)$~\cite{BELLEsuppl}. We choose the third one because it contains the most complete information about the $(z, M_h)$ dependence of DiFFs, including their correlations (see 
Sec.~\ref{sec:D1fit}). As reported in Tab.~VIII of Ref.~\cite{BELLEsuppl}, only 58 of the 64 bins are filled. We use 46 of them by dropping the highest bin in $z$ ($[0.8, 1]$) and in $M_h$ 
($[1.5, 2.0]$) because they are scarcely populated and our description of $D_1$ is worse. The upper cut in $M_h$ is also consistent with the grid used in the Monte Carlo simulation of the unpolarized cross section (see Sec.~\ref{sec:D1MC}). 

Using {\tt MINUIT}, we minimize 
\begin{equation}
\chi^2 = \sum_{ij}  \frac{\left( H_{ij}^{\mathrm{th}} - H_{ij}^{\mathrm{exp}} \right)^2}{\sigma_{ij}^2} \; , 
\label{eq:chi2A}
\end{equation}
where $H_{ij}^{\mathrm{exp}}$ is obtained using Eq.~\eqref{eq:ACasyfit}. Namely, for each bin 
$(z_i, \, M_{h\, j})$ the average value of angles $\theta_2,\, \theta,$ and $\overline{\theta}$, is taken from Ref.~\cite{BELLEsuppl}. Then, using Eqs.~\eqref{eq:ACasyden} and~\eqref{eq:dsig0-MC-bin} the contribution $D_{ij}$ of the function $D$ is defined as 
\begin{equation}
\begin{split}
D_{ij} &\equiv \int_{z_i}^{z_i+\Delta z} dz \int_{M_{h\, j}}^{M_{h\, j}+\Delta M_h} dM_h \, 
D(z, M_h; Q^2) \\
&= \frac{1}{4\pi \alpha^2 / Q^2} \, \sum_{q=u,d,s,c} \, n_q (Q^2) \, \sum_{\mathrm{ch}} \, 
( d\sigma^{0\, q}_{\mathrm{ch}} )_{ij} \; , 
\label{eq:Dgrid}
\end{split}
\end{equation}
where $( d\sigma^{0\, q}_{\mathrm{ch}} )_{ij}$ fits the Monte Carlo simulation of the unpolarized cross section for the considered bin, channel ch, and flavor $q$. 
By summing the latter over all experimental bins and channels (and dividing by the factor 
$4\pi \alpha^2 / Q^2$), we get the $n_q (Q^2)$ for each flavor. Finally, in Eq.~\eqref{eq:ACasyfit} the Artru--Collins asymmetry $A$ for the bin $(z_i, \, M_{h\, j})$ is taken from the Belle 
measurement~\cite{Vossen:2011fk}.


The error $\sigma_{ij}$ in Eq.~\eqref{eq:chi2A} is obtained by summing the statistical and systematic errors in quadrature for the measurement of $A$ reported by the Belle collaboration~\cite{BELLEsuppl}, multiplied by all factors relating $A$ to $H$ according to Eq.~\eqref{eq:ACasyfit}. The sum runs upon the above mentioned 46 bins. 



The last ingredient of the $\chi^2$ formula is $H_{ij}^{\mathrm{th}}$. It is obtained by first parametrizing the function $H$ in Eq.~\eqref{eq:ACasyfit} at the starting scale $Q_0^2$ as
\begin{equation}
\begin{split}
H(z, M_h; Q_0^2) &= N 2 |\bm{R}|\, (1-z) \, \exp [\gamma_1 (z - \gamma_2 M_h)] \\
&\mbox{\hspace{-1cm}} \times \Bigg[ P(0, 1, \delta_1, 0, 0 ; z) + 
z P(0, 0, \delta_2, \delta_3, 0; M_h) \\
&\mbox{\hspace{-0.5cm}} + \frac{1}{z} \, P(0, 0, \delta_4, \delta_5, 0; M_h) \Bigg] \, 
\mathrm{BW} \left( m_\rho, \frac{\eta}{m_\rho} ; M_h \right) \; , 
\label{eq:Hfit}
\end{split}
\end{equation}
where the polynomial $P$ and the function BW are defined in Eq.~\eqref{eq:struct}~\footnote{Note that 
Eq.~\eqref{eq:struct} is proportional to the modulus squared of a relativistic Breit--Wigner, but also to its imaginary part. Therefore, the parametrization in Eq.~\eqref{eq:Hfit} is in agreement with the assumption that $H_{1, sp}^{\open q}$ is given by the interference between a relative $s$ wave and a relative $p$ 
wave~\cite{Bacchetta:2006un}}. Then, we evolve it at the Belle scale $Q^2$ using the {\tt HOPPET} 
code~\cite{Salam:2008qg}, suitably extended to include LO chiral-odd splitting functions, and we integrate it on the considered bin $(z_i, \, M_{h\, j})$. 

By minimizing the $\chi^2$ of Eq.~\eqref{eq:chi2A}, we get the best values for the 9 parameters 
$N,\, \gamma_{i=1,2},\, \delta_{i=1-5},\, \eta$. They are listed in Tab.~\ref{tab:params}, together with their statistical errors obtained from the condition $\Delta \chi^2 = 1$. The $\chi^2 /$ dof turns out to be 0.57.

\begin{table}[h]
\begin{tabular}{|c|c|}
\hline
$N=0.0132\pm 0.0033$ &   \\
\hline
$\gamma_1=-2.873\pm 0.229 $ & $\gamma_2=-0.644\pm 0.094$  \\
\hline
$\delta_1=23.310\pm 7.534$ & $\delta_2=-199.410\pm 17.728$ \\
\hline
 $\delta_3=276.920\pm 20.511$ &  $\delta_4=36.732\pm 3.796$ \\
 \hline
 $\delta_5=-42.406\pm 4.427$ & $\eta=0.303\pm 0.023$  \\
\hline
\end{tabular}
\caption{The free parameters with their statistical errors from Eq.~\eqref{eq:Hfit}, obtained by fitting the experimental Artru--Collins asymmetry of Ref.~\cite{Vossen:2011fk}.} 
\label{tab:params} 
\end{table} 


By summing $H (z_i, M_{h\, j}; Q^2)$ over all bins, we get the $[ n_u^\uparrow (Q^2)]^2$ of 
Eq.~\eqref{eq:Hnorm}. In the last step, we get the polarized DiFF $H_{1,\, sp}^{\open\, u}$ bin by bin from Eq.~\eqref{eq:H1}.

\begin{figure}[h]
\begin{center}
\includegraphics[width=9cm]{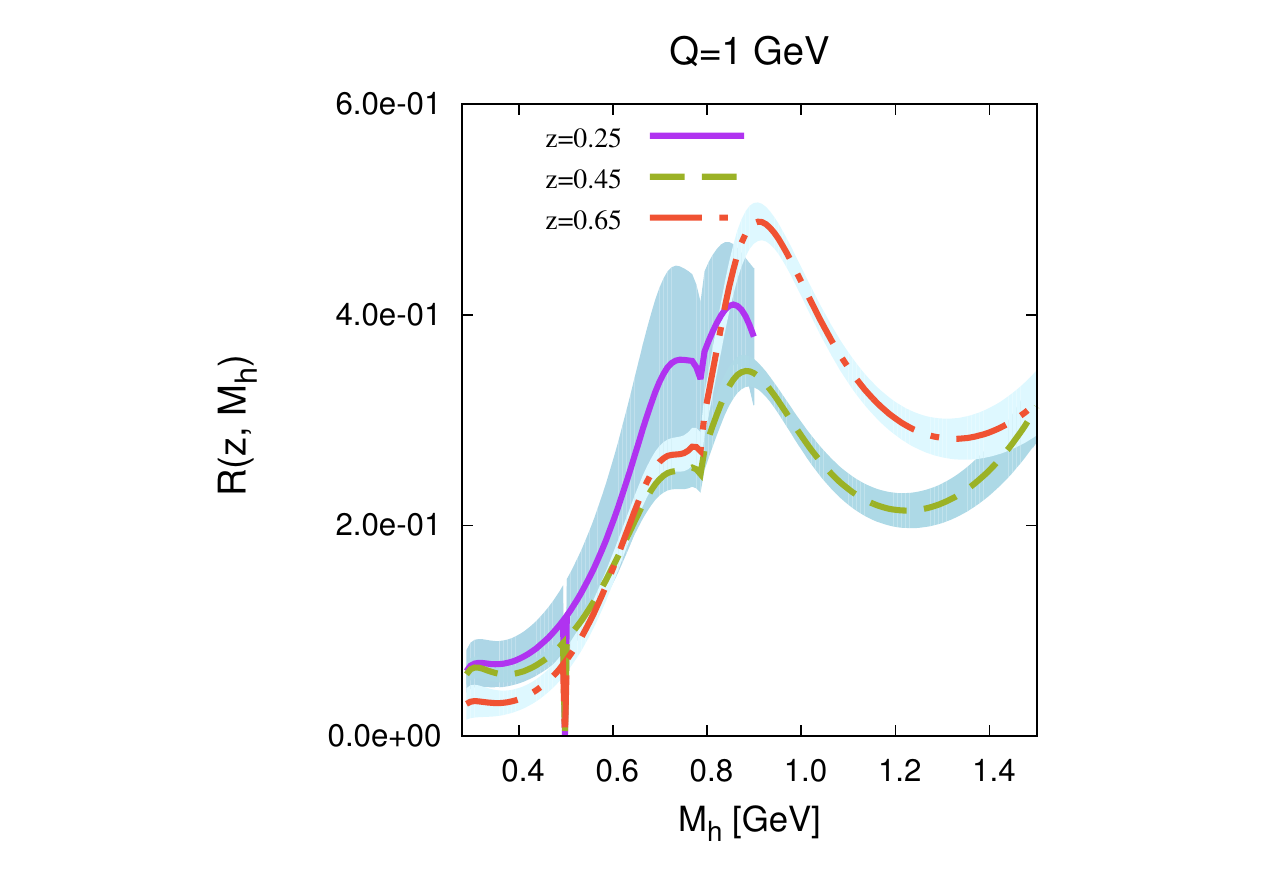}\\
\includegraphics[width=9cm]{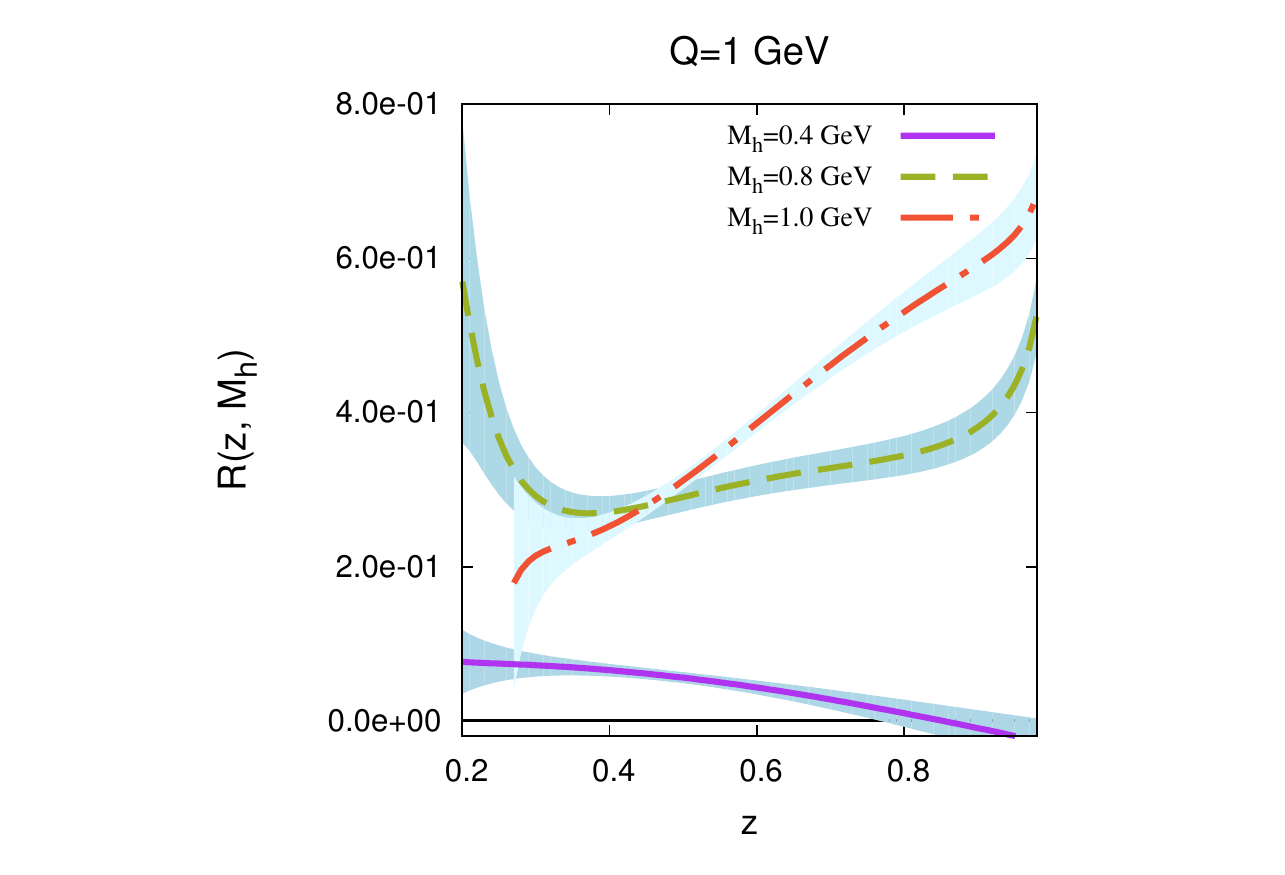}
\end{center}
\vspace{-0.5cm}
\caption{The ratio $R$ of Eq.~\eqref{eq:Ratio}, summed over all channels, at the hadronic scale 
$Q_0^2 = 1$ GeV$^2$. Upper panel for $R$ as a function of $M_h$ for $z=0.25$ (solid line), $z=0.45$ (dashed line), and $z=0.65$ (dot-dashed line). Lower panel for $R$ as a function of $z$ for $M_h=0.4$ GeV (solid line), $M_h=0.8$ GeV (dashed line), and $M_h=1.0$ GeV (dot-dashed line). For the calculation of the uncertainty bands, see details in the text. The ratio is affected also by a 10\% systematic error.}
\label{fig:R}
\end{figure}


\subsection{Results for $H_1^{\open}$}
\label{sec:H1results}

In Fig.~\ref{fig:R}, we show the ratio
\begin{equation}
R(z, M_h) = \frac{|\bm{R}|}{M_h} \, \frac{H_{1, sp}^{\open u}(z, M_h; Q_0^2)}
                                                                      {D_1^u (z, M_h; Q_0^2)} \; , 
\label{eq:Ratio}
\end{equation}
summed over all channel, at the hadronic scale $Q_0^2 = 1$ GeV$^2$. The upper panel displays the ratio as a function of $M_h$ at three values of $z$: 0.25 (solid line), 0.45 (dashed line), and 0.65 (dot-dashed line). The lower panel displays it as a function of $z$ at $M_h=0.4$ GeV (solid line), 0.8 GeV (dashed line), and 1 GeV (dot-dashed line). The uncertainty bands correspond to the statistical errors of the fitting parameters (see Tab.~\ref{tab:params}). They are calculated through the standard procedure of error propagation using the covariance matrix provided by {\tt MINUIT} (with $\Delta \chi^2 = 1$). Due to differences between the Monte Carlo simulation and the experimental cross section, we estimated a 10\% systematic error in the determination of $R$. In the upper panel, the solid line stops at $M_h=0.9$ GeV because there are no experimental data at higher invariant masses for $z=0.25$. The fit is less constrained in that region and the error band becomes larger. The same effect is visible in the lower panel for the highest displayed $M_h$ (dot-dashed line) at low $z$. Note that in the upper panel all three curves display a dip at $M_h \sim 0.5$ GeV. It corresponds to the peak for the $K^0_S \to \pi^+ \pi^-$ decay, which is present in the denominator of $R$ (via $D_1^u$) but not in the numerator (we recall that we assume $H_{1, sp}^{\open u} \approx 0$ for this channel, see the discussion after Eqs.~\eqref{eq:ass3} and~\eqref{eq:ass4}).

\begin{figure}[h]
\begin{center}
\includegraphics[width=8.2cm]{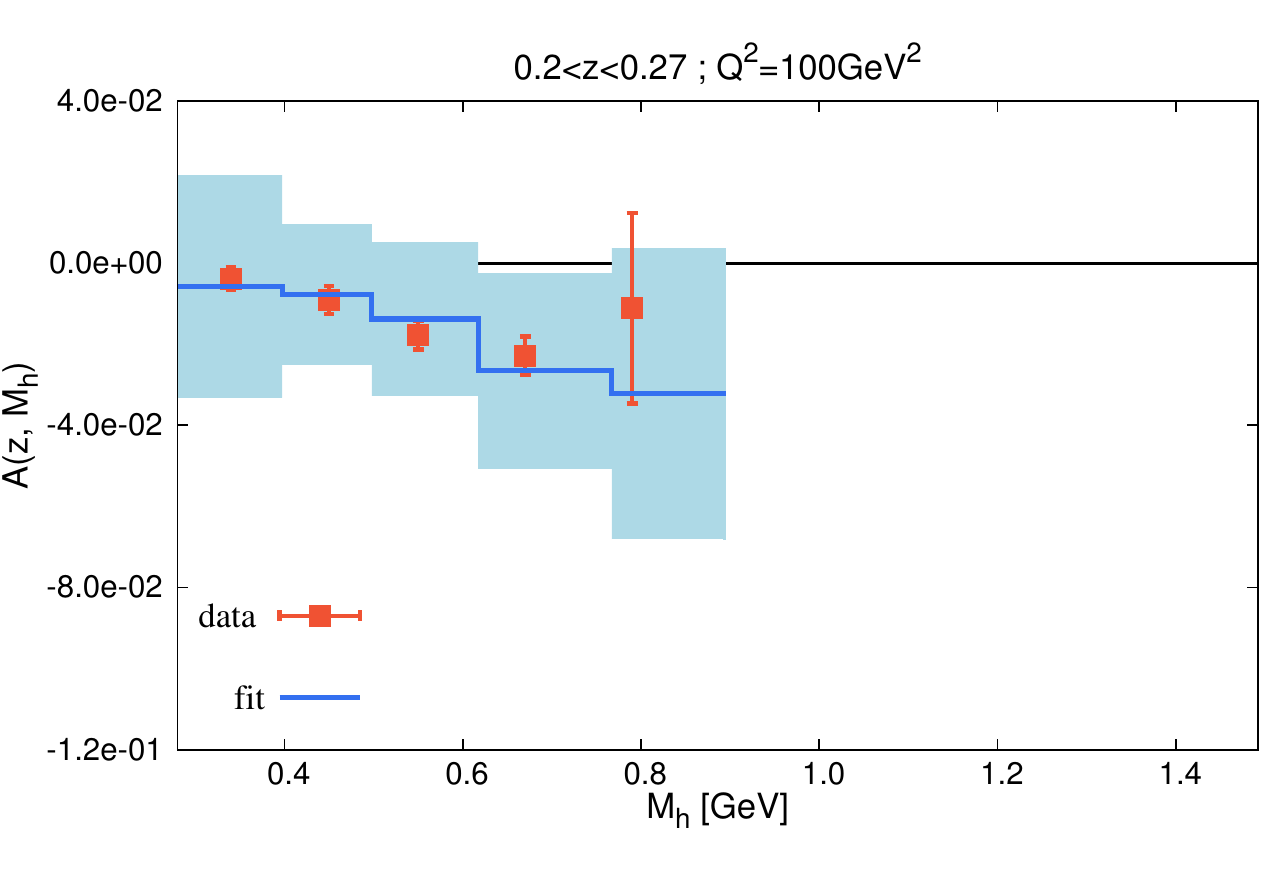}\hspace{0.5cm}\includegraphics[width=8.2cm]{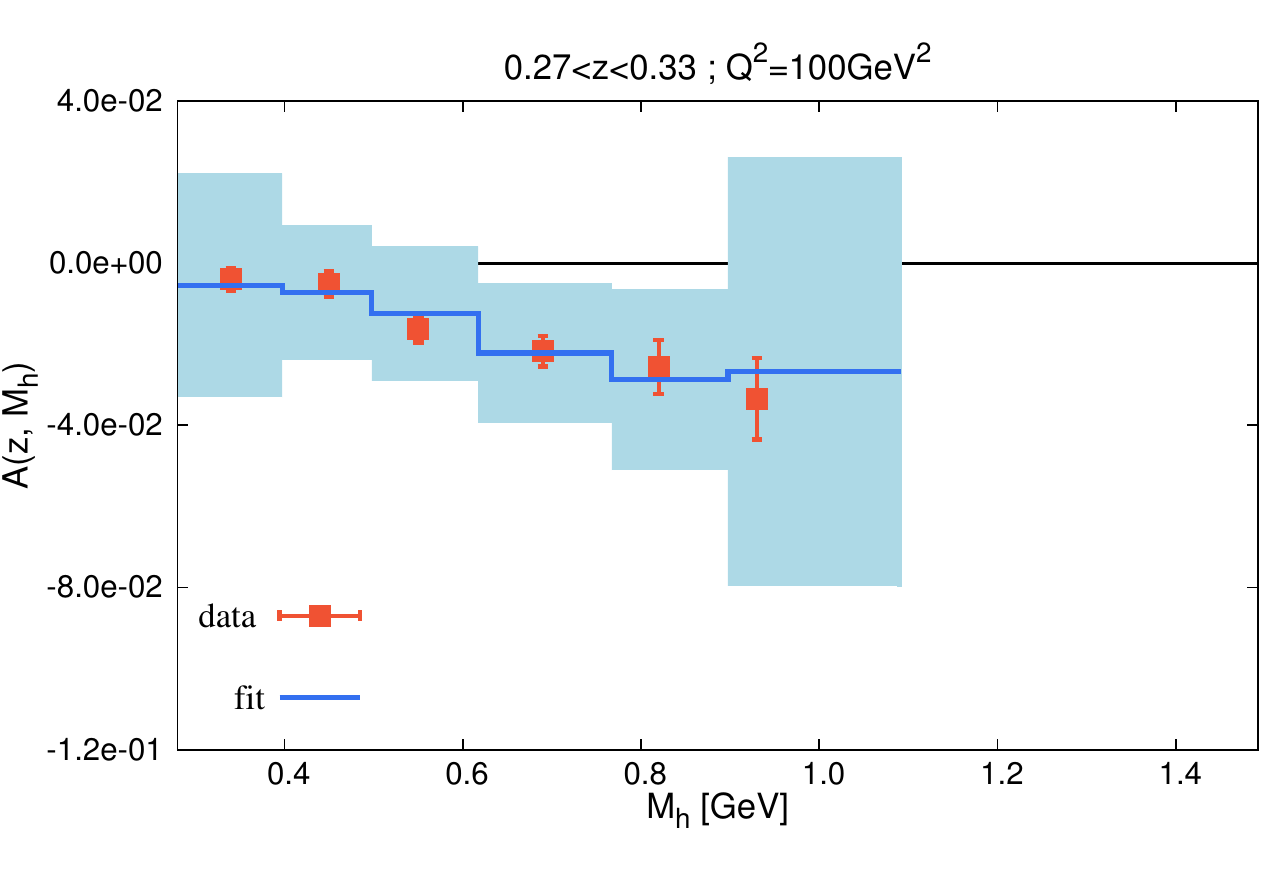}\\
\includegraphics[width=8.2cm]{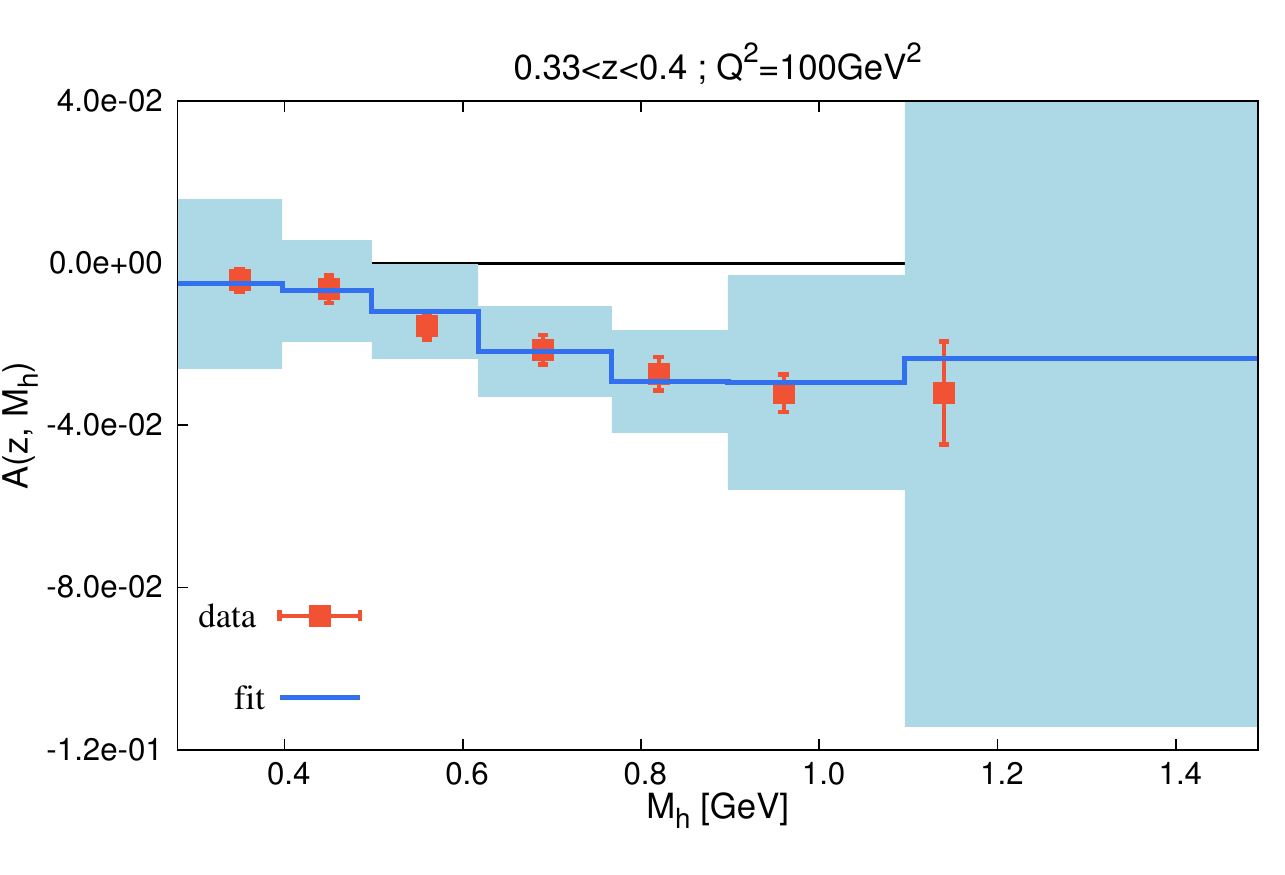}\hspace{0.5cm}\includegraphics[width=8.2cm]{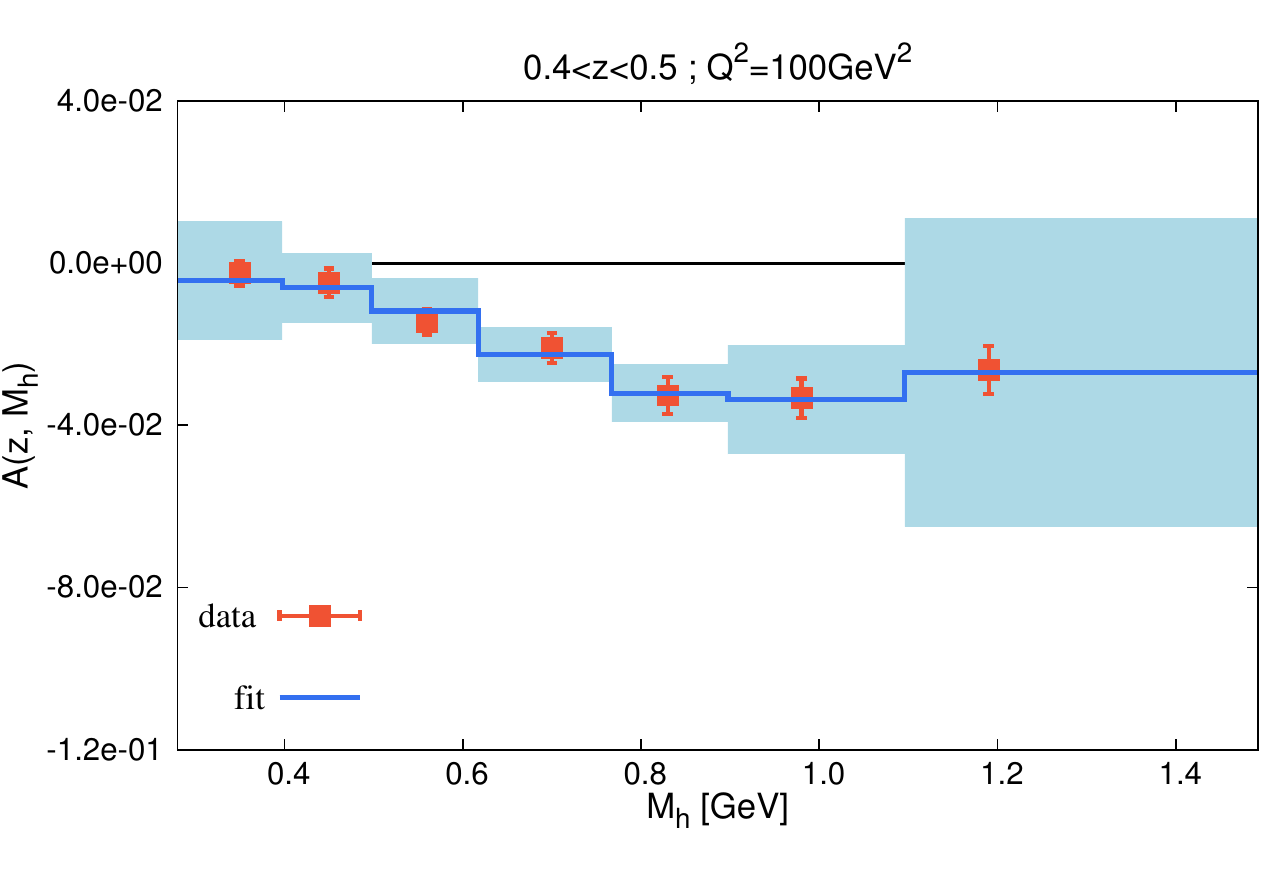}\\
\includegraphics[width=8.2cm]{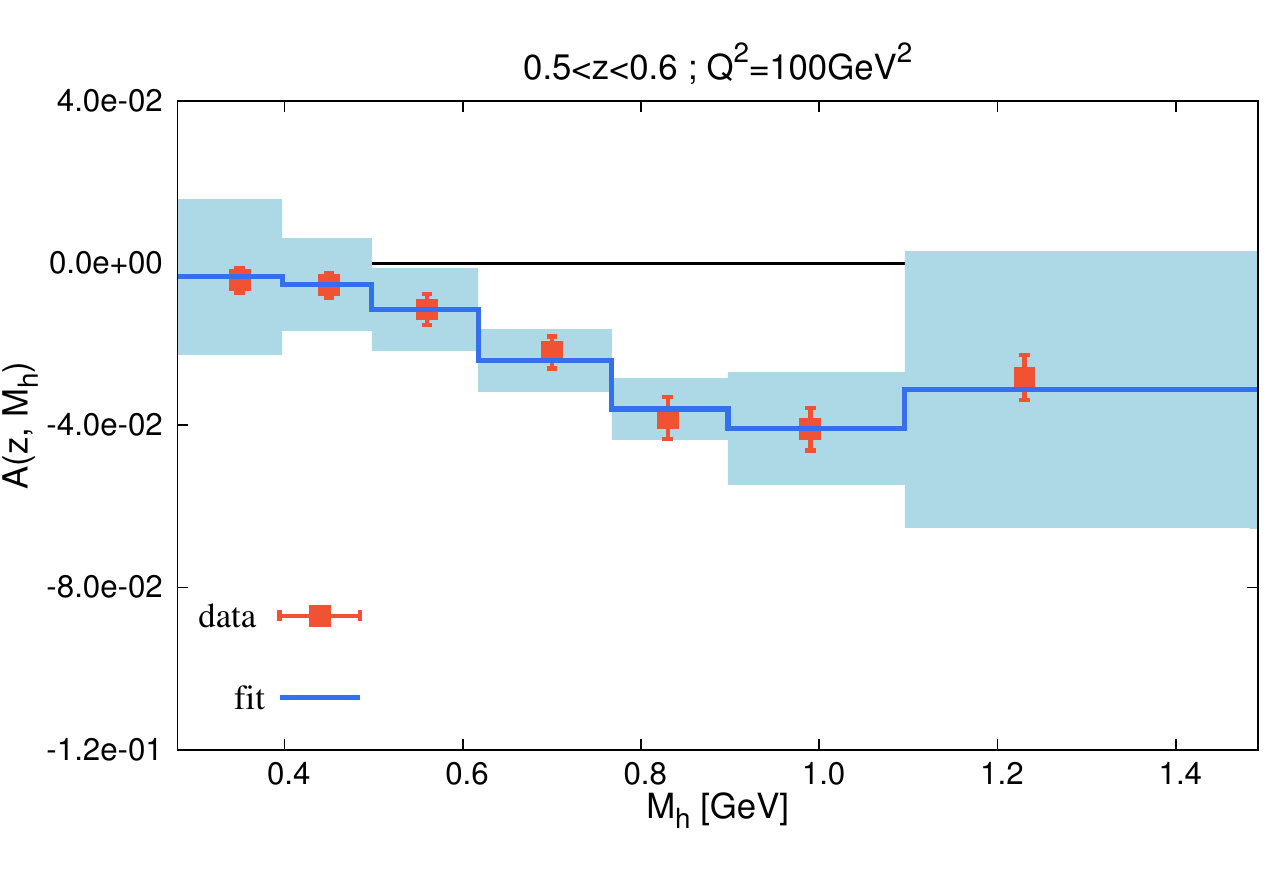}\hspace{0.5cm}\includegraphics[width=8.2cm]{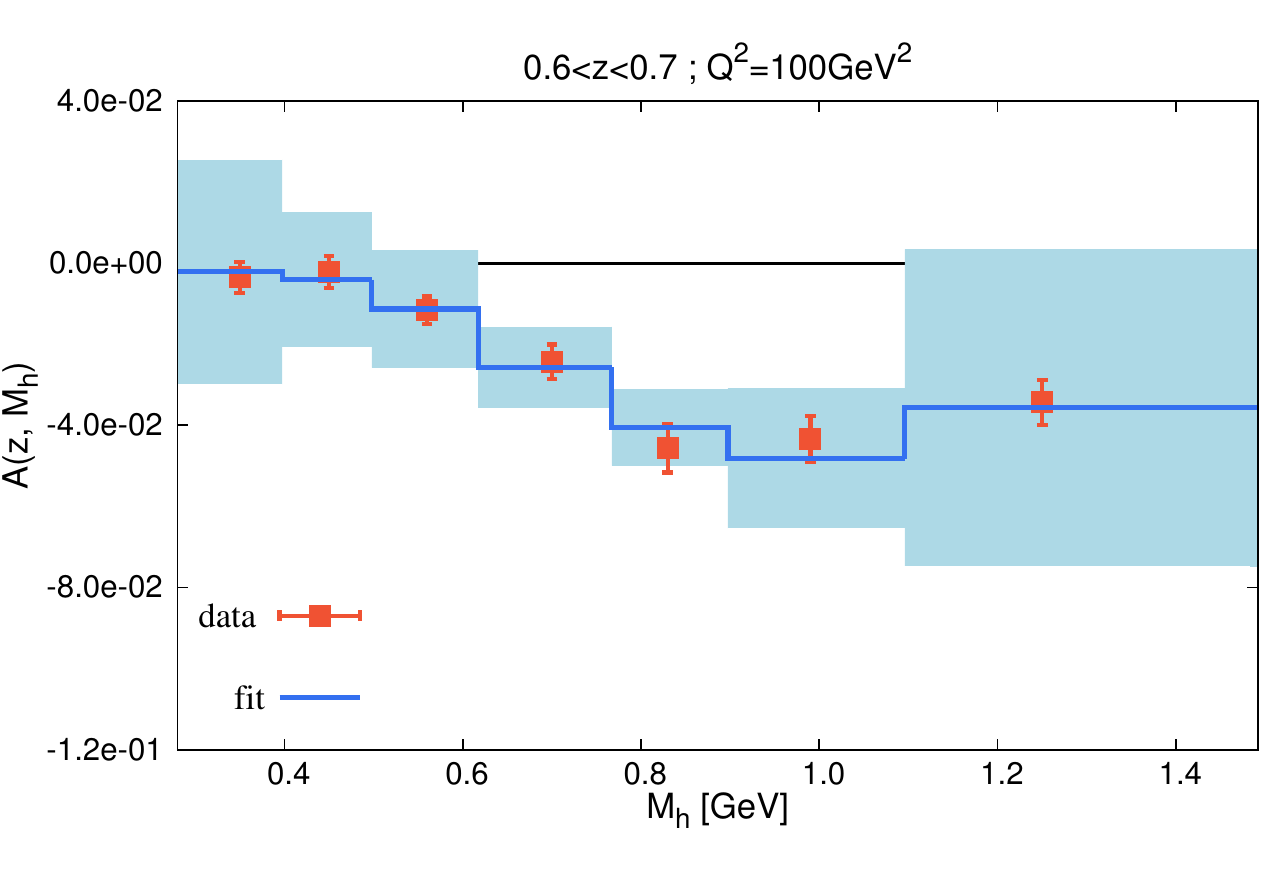}\\
\includegraphics[width=8.2cm]{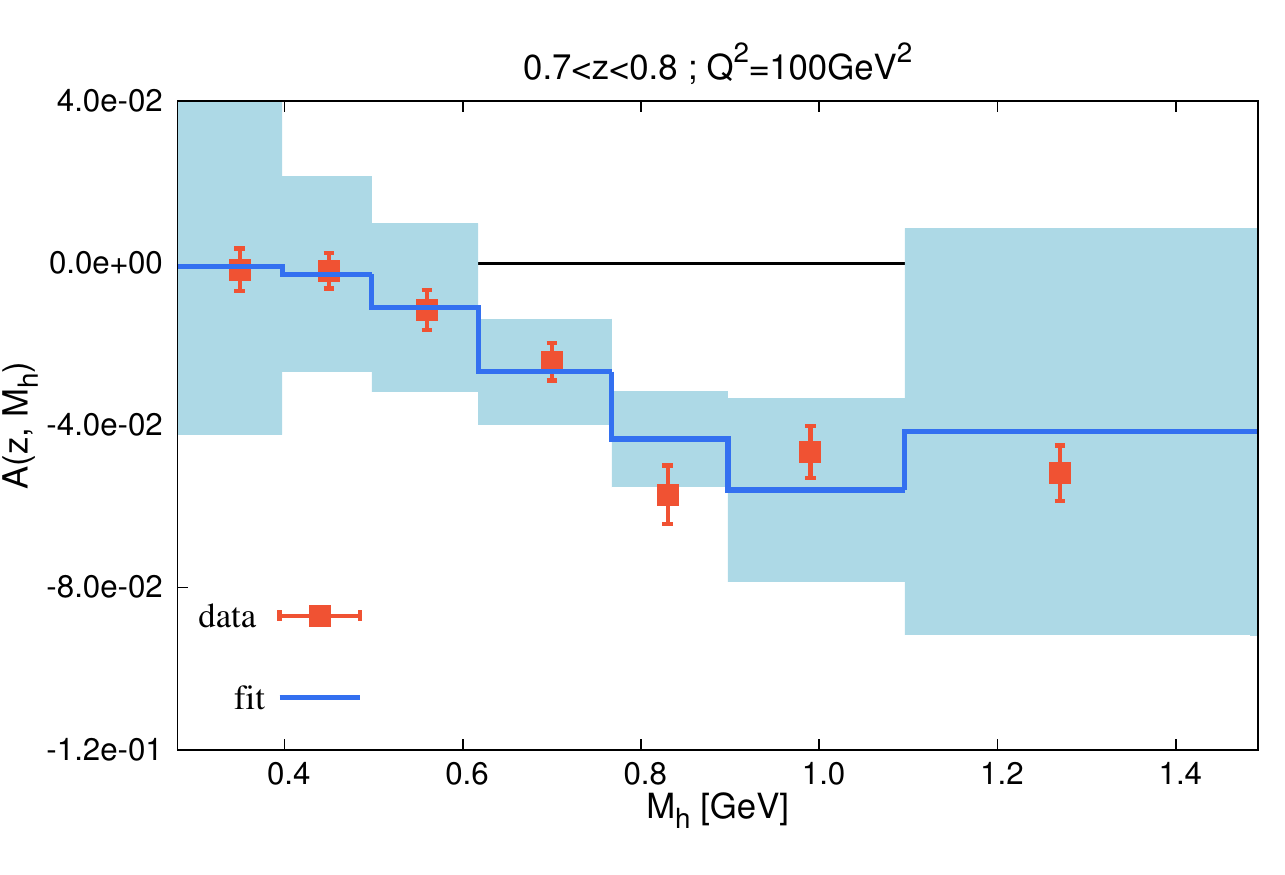}
\end{center}
\vspace{-0.5cm}
\caption{The Artru--Collins asymmetry at $Q^2=100$ GeV$^2$ for the experimental bins 
$(z_i, M_{h\, j})$. Points with error bars for the measurement by Belle~\cite{Vossen:2011fk}. The solid line represents the top side of the histogram for the fitting formula of Eq.~\eqref{eq:Agrid}. The shaded area is the corresponding statistical error (see text for more details).}
\label{fig:Aexp}
\end{figure}

In Fig.~\ref{fig:Aexp}, we show the Artru--Collins asymmetry at $Q^2=100 $ GeV$^2$. Each panel corresponds to the indicated experimental $z$ bin, ranging from $[0.2, 0.27]$ to $[0.7, 0.8]$. In each panel, the points with error bars indicate the Belle measurement for the experimental 
$M_h$ bins~\cite{BELLEsuppl}. For each bin $(z_i, M_{h\, j})$, the solid line represents the top side of the histogram for the fitting asymmetry obtained by inverting Eq.~\eqref{eq:ACasyfit}, i.e.
\begin{equation}
A_{ij}^{\mathrm{th}} = - \frac{\langle \sin^2 \theta_2 \rangle}{\langle 1+\cos^2 \theta_2\rangle}
           \,\langle \sin\theta \rangle\,\langle \sin\overline{\theta}\rangle\, \frac{5}{9} \, 
           \frac{H_{ij}^{\mathrm{th}}}{D_{ij}} \; , 
\label{eq:Agrid}
\end{equation}
where $D_{ij}$ is defined in Eq.~\eqref{eq:Dgrid}, $H_{ij}^{\mathrm{th}}$ is defined in the discussion about Eq.~\eqref{eq:Hfit}, and the average values of the angles in the considered bin are taken from Ref.~\cite{BELLEsuppl}. The shaded areas are the statistical errors of 
$A_{ij}^{\mathrm{th}}$, deduced from the parameter errors in Tab.~\ref{tab:params} through the standard formula for error propagation. Note that the statistical uncertainty of the fit is very large for the highest $M_h$ bin.


\section{Conclusions and Outlooks}
\label{sec:end}

In this paper, we have parametrized for the first time the full dependence of the dihadron fragmentation functions (DiFFs) that describe the nonperturbative fragmentation of a hard parton into two hadrons inside the same jet, plus other unobserved fragments. The dependence of DiFFs on the invariant mass and on the energy fraction carried by a $(\pi^+ \pi^-)$ pair produced in $e^+ e^-$ annihilations, is extracted by fitting the recent Belle data~\cite{Vossen:2011fk}. 

The analytic formulae for both unpolarized and polarized DiFFs at a starting hadronic scale are inspired by previous model calculations of 
DiFFs~\cite{Bianconi:1999uc,Bacchetta:2006un,Bacchetta:2008wb}. Then, they are evolved at leading order using the {\tt HOPPET} code~\cite{Salam:2008qg}, suitably extended to include chiral-odd splitting functions that can describe scaling violations of  chiral-odd polarized DiFFs. 

In the absence of published data for the unpolarized cross section, we extract the unpolarized DiFF (appearing in the denominator of the asymmetry)  by fitting the simulation produced by the 
{\tt PYTHIA} event generator~\cite{Sjostrand:2003wg} at Belle kinematics, since this code is known to give a good description of the $e^+ e^-$ total cross section~\cite{courtesyBELLE}. 

Given the rich structure of the invariant mass distribution in the selected range $[2m_\pi, \, 1.3]$ GeV, we have considered three different channels for producing a $(\pi^+ \pi^-)$ pair (via 
$\rho, \, \omega,$ or $K^0_S$ decays), as well as a continuum channel that includes everything 
else~\cite{Bacchetta:2006un}. The analysis is performed at leading order; gluons are generated only radiatively. In the Monte Carlo simulation of the unpolarized cross section, more than 1 million 
$(\pi^+ \pi^-)$ pairs are collected in 31585 bins and their distribution is fitted using {\tt MINUIT}, reaching a global $\chi^2$/dof of 1.62. Statistical errors are small because of the large statistics in the Monte Carlo. Experimental data for the Artru--Collins asymmetry are collected instead in 46 bins and are fitted with a 
9-parameters function getting a final $\chi^2$/dof of 0.57. 

The long-term goal of this work is to improve the above analysis by repeating the Monte Carlo simulation at different hard scales. In this way, we should be able to better constrain the evolution of the unpolarized DiFF and to reduce the systematic uncertainty deriving from the arbritariness in the choice of the analytic expression at the starting hadronic scale. Moreover, including also data with asymmetries for $(\pi , K)$ and $(K, K)$ pairs the flavor analysis would improve beyond the present limitations induced by isospin symmetry and charge conjugation applied to $(\pi^+ \pi^-)$ pairs only. 

As we make progress in the knowledge of DiFFs, it is crucial to have new data on hadron pair production officially released. Using the COMPASS data on semi-inclusive deep-inelastic scattering on transversely polarized protons and deuterons~\cite{Wollny:2010}, we will be able to update the results of Ref.~\cite{Bacchetta:2011ip} about  the extraction of the transversity parton distribution. From the PHENIX data on (polarized) proton-proton collisions~\cite{Yang:2009zzr}, we can also explore an alternative extraction of transversity~\cite{Bacchetta:2004it}, aiming at studying the yet unknown contribution from antiquarks. 


\widetext

\section*{Acknowledgments}

We thank the Belle collaboration, particularly A.~Vossen, for several enlightening discussion about the experimental analysis, and for making available the details about the Monte Carlo simulation of the unpolarized cross section. A.~Courtoy is presently working under the Belgian Fund F.R.S.-FNRS via the contract of Charg\'ee de recherches. This work is partially supported by the Italian MIUR through the PRIN 2008EKLACK, and by the Joint Research Activity ``Study of Strongly Interacting Matter" (acronym HadronPhysics3, Grant Agreement No. 283286) under the 7th Framework Programme of the European Community. 


\appendix


\section{Functional form of $D_1$ at $Q_0^2 = 1$ GeV$^2$}
\label{sec:A}

In this appendix, we list the analytic formulae for the unpolarized DiFF $D_{1,\, \mathrm{ch}}^q$ at the hadronic scale $Q_0^2 = 1$ GeV$^2$ for each flavor $q = u,\, d,\, s,\, c,$ and for the resonant channels $\rho, \, \omega,$ and $K^0_S$, as well as for the continuum. For each case, we add a table with the best-fit values and statistical errors of the involved parameters. 

We recall that the recurring structures of the polynomial $P (a_1, a_2, a_3, a_4, a_5; x)$ and the function 
$\mathrm{BW} (m, \Gamma; x)$ are defined in Eq.~\eqref{eq:struct}. 

\vspace{0.5cm}
\subsection{Functional form of the continuum channel at $Q_0^2 = 1$ GeV$^2$}
\label{sec:Acont}

\subsubsection{up and down}
\vspace{-0.2cm}
\begin{equation}
\begin{split}
&D_{1,\mathrm{cont}}^u (z, M_h; Q_0^2) = N_1^c \, z^{\alpha_1^c} (1-z)^{(\alpha_2^c)^2} 
(2 |\bm{R}| )^{(\beta_1^c)^2}\,
\exp \left[ - \left( P(\gamma_1^c, \gamma_2^c, \gamma_3^c, 0, 0; z) + \frac{\gamma_4^c}{M_h} 
\right)^2 \, ( 2 |\bm{R}| )^2 \right]   \\
&D_{1,\mathrm{cont}}^d (z, M_h; Q_0^2) = D_{1,\mathrm{cont}}^u (z, M_h; Q_0^2)  \; , 
\label{eq:cont-ud}
\end{split}
\end{equation}

with best-fit parameters

\begin{table}[h]
\begin{tabular}{|c||c|c|}
\hline
cont &  &   \\
\hline
\hline
$u=d$ & $N_1^c=0.601\pm 0.013$ & $\beta_1^c =0.8446\pm 0.0059$ \\
\hline
&  $\alpha_1^c =-2.282\pm 0.018$ & $\alpha_2^c =1.0012\pm 0.0072$  \\
\hline
& $\gamma_1^c =0.7133\pm 0.0083$ & $\gamma_2^c =-0.155\pm 0.038$  \\
\hline
&  $\gamma_3^c =1.180\pm 0.044$ & $\gamma_4^c =-1.051\pm 0.017$  \\
\hline
\end{tabular}
\caption{} 
\label{tab:cont-ud} 
\end{table} 
\vspace{0.5cm}

\subsubsection{strange}
\begin{equation}
D_{1,\mathrm{cont}}^s (z, M_h; Q_0^2) = (N_2^c)^2 (1-z)^{(\alpha_3^c)^2} \,  
D_{1,\mathrm{cont}}^u (z, M_h; Q_0^2)  \; , 
\label{eq:cont-s}
\end{equation}
with best-fit parameters
\begin{table}[h]
\begin{tabular}{|c||c|c|}
\hline
cont &  &  \\
\hline
\hline
$s$ & $N_2^c=0.7825\pm 0.0038$ & $\alpha_3^c =0.636\pm 0.012$  \\
\hline
\end{tabular}
\caption{} 
\label{tab:cont-s} 
\end{table} 

\subsubsection{charm}
\begin{equation}
D_{1,\mathrm{cont}}^c (z, M_h; Q_0^2) = N_3^c \, z^{\alpha_4^c} \, (1-z)^{(\alpha_5^c)^2} 
(2 |\bm{R}| )^{(\beta_2^c)^2}  \, 
\exp \left[ - \left( P(\gamma_5^c, 0, \gamma_6^c, 0, 0; z) + \frac{\gamma_7^c}{M_h} \right)^2 \, 
( 2 |\bm{R}| )^2 \right]    \; , 
\label{eq:cont-c}
\end{equation}
with best-fit parameters
\begin{table}[h]
\begin{tabular}{|c||c|c|}
\hline
cont &  &   \\
\hline
\hline
$c$ & $N_3^c=1.437\pm 0.054$ & $\beta_2^c =0.940\pm 0.010$ \\
\hline
& $\alpha_4^c =-2.310\pm 0.027$ & $\alpha_5^c =1.7020\pm 0.0080$  \\
\hline
 & $\gamma_5^c =0.6336\pm 0.0059$ & $\gamma_6^c =0.816\pm 0.018$  \\
\hline
&  $\gamma_7^c =-0.645\pm 0.030$ &   \\
\hline
\end{tabular}
\caption{} 
\label{tab:cont-c} 
\end{table} 


\subsection{Functional form of the $\rho$ channel at $Q_0^2 = 1$ GeV$^2$}
\label{sec:Arho}

\subsubsection{up and down}

The $D_{1, \rho}^u (z, M_h; Q_0^2)$ is defined in Eq.~\eqref{eq:rho-u} and the best values of its parameters are reported in Tab.~\ref{tab:rho-ud}. Then, we take 
\begin{equation}
D_{1, \rho}^d (z, M_h; Q_0^2) = D_{1, \rho}^u (z, M_h; Q_0^2)  \; . 
\label{eq:rho-d}
\end{equation}

\subsubsection{strange}
\vspace{-0.5cm}
\begin{equation}
\begin{split}
D_{1, \rho}^s (z, M_h; Q_0^2) &= (N_2^\rho)^2 z^{\alpha_3^\rho} (1-z)  \, 
D_{1, \rho}^u (z, M_h; Q_0^2)  \; , 
\label{eq:rho-s}
\end{split}
\end{equation}

with best-fit parameters

\begin{table}[h]
\begin{tabular}{|c||c|c|}
\hline
$\rho$ &  &  \\
\hline
\hline
$s$ & $N_2^\rho  = 0.861\pm 0.074$ & $\alpha_3^\rho = -0.244\pm 0.110$  \\
\hline
\end{tabular}
\caption{} 
\label{tab:rho-s} 
\end{table} 

\subsubsection{charm}
\vspace{-0.5cm}
\begin{equation}
\begin{split}
D_{1, \rho}^c (z, M_h; Q_0^2) &= (N_3^\rho)^2 z^{\alpha_4^\rho} (1-z)^{(\alpha_5^\rho)^2} 
(2 |\bm{R}| )^{(\beta_2^\rho)^2}  \\
&\times \Bigg[ 
   \exp \left[ - P(0, \gamma_4^\rho, 0, 0, -\gamma_4^\rho; z)\, M_h^2  \right]   \, 
    \exp \left[ - P(\delta_3^\rho, 0, \delta_4^\rho, 0, 0; z) \right]  + 
    (\eta_2^\rho )^2 \, \mathrm{BW} (m_\rho, \Gamma_\rho; M_h) \Bigg]  \; , 
\label{eq:rho-c}
\end{split}
\end{equation}
with best-fit parameters
\begin{table}[h]
\begin{tabular}{|c||c|c|}
\hline
$\rho$ &  &  \\
\hline
\hline
$c$ & $N_3^\rho  = 0.450\pm 0.031$ & $\beta_2^\rho = 0.697\pm 0.028$ \\
\hline
& $\alpha_4^\rho = 1.850\pm 0.093$ & $\alpha_5^\rho = 2.474\pm 0.025$  \\
\hline
 & $\gamma_4^\rho = 3.958\pm 0.357$ &  $\eta_2^\rho = 2.223\pm 0.081$ \\
\hline
 & $\delta_3^\rho = -1.220 \pm 0.066$  &  $\delta_4^\rho = 3.721 \pm 1.234$  \\ 
\hline
\end{tabular}
\caption{} 
\label{tab:rho-c} 
\end{table} 

\vspace{0.5cm}

\subsection{Functional form of the $\omega$ channel at $Q_0^2 = 1$ GeV$^2$}
\label{sec:Aomega}

\subsubsection{up and down}
\vspace{-0.5cm}
\begin{equation}
\begin{split}
D_{1, \omega}^u (z, M_h; Q_0^2) &= (1-z)^{(\alpha_1^\omega)^2} \, 
(2 |\bm{R}| )^{\beta_1^\omega} \,  \frac{1}{1+\exp[5 (M_h-1.2)]} \\
&\times \Bigg[  N_1^\omega \, 
   \exp \left[ - P(\gamma_1^\omega, \gamma_2^\omega, \gamma_3^\omega, 0, 0; z)\, 
                      (2 |\bm{R}|)^{2\beta_2^\omega}  \right]   \,
    \exp \left[ - P(\delta_1^\omega, 0, \delta_2^\omega, 0, 0; z) \right]  + 
    (\eta_1^\omega)^2\, \mathrm{BW} (m_\omega, \Gamma_\omega; M_h) \Bigg]  \\
D_{1, \omega}^d (z, M_h; Q_0^2) &= D_{1, \omega}^u (z, M_h; Q_0^2)  \; , 
\label{eq:omega-ud}
\end{split}
\end{equation}
with $m_\omega = 0.783$ GeV and $\Gamma_\omega = 0.008$ GeV, and with best-fit parameters
\begin{table}[h]
\begin{tabular}{|c||c|c|}
\hline
$\omega$ &  &   \\
\hline
\hline
$u=d$ & $N_1^\omega  = 3.234 \times 10^{14}\pm 4.377 \times 10^{13}$ & 
$\alpha_1^\omega = 1.220\pm 0.025$ \\
\hline
& $\beta_1^\omega = 12.539\pm 0.083$ &  $\beta_2^\omega = 0.2899\pm 0.0019$  \\
\hline
 & $\gamma_1^\omega = 1.970\pm 0.105$ & $\gamma_2^\omega = 31.032\pm 0.328$ \\
 \hline
 & $\gamma_3^\omega = 10.228\pm 0.736$ & $\eta_1^\omega = 0.0388 \pm 0.0010$  \\
\hline
 & $\delta_1^\omega = -0.862 \pm 0.061$  &  $\delta_2^\omega = -0.279 \pm 0.445$  \\ 
\hline
\end{tabular}
\caption{} 
\label{tab:omega-ud} 
\end{table} 

\subsubsection{strange}
\vspace{-0.5cm}
\begin{equation}
\begin{split}
D_{1, \omega}^s (z, M_h; Q_0^2) &= (N_2^\omega)^2 z^{\alpha_2^\omega}  \, 
D_{1, \omega}^u (z, M_h; Q_0^2)  \; , 
\label{eq:omega-s}
\end{split}
\end{equation}

with best-fit parameters

\begin{table}[h]
\begin{tabular}{|c||c|c|}
\hline
$\omega$ &  &   \\
\hline
\hline
$s$ & $N_2^\omega  = 0.297\pm 0.010$ & $\alpha_2^\omega = -1.233\pm 0.058$  \\
\hline
\end{tabular}
\caption{} 
\label{tab:omega-s} 
\end{table} 

\subsubsection{charm}
\vspace{-0.5cm}
\begin{equation}
\begin{split}
D_{1, \omega}^c (z, M_h; Q_0^2) &=  (1-z)^{(\alpha_3^\omega)^2} 
(2 |\bm{R}| )^{\beta_3^\omega} \, \frac{1}{1+\exp [5 (M_h-1.2)]} \\
&\times \Bigg[  N_3^\omega \, 
   \exp \left[ - P(0, \gamma_4^\omega, 0, 0, 0; z)\, 
                      (2 |\bm{R}|)^{2\beta_4^\omega}  \right]  \, 
   \exp \left[ - P(\delta_3^\omega, 0, \delta_4^\omega, 0, 0; z) \right]  + 
   (\eta_2^\omega)^2\, \mathrm{BW} (m_\omega, \Gamma_\omega; M_h) \Bigg]  \; , 
\label{eq:omega-c}
\end{split}
\end{equation}
with best-fit parameters
\begin{table}[h]
\begin{tabular}{|c||c|c|}
\hline
$\omega$ &  &  \\
\hline
\hline
$c$ & $N_3^\omega  = 1.758 \times 10^{13}\pm 2.428 \times 10^{12}$ & 
$\alpha_3^\omega = 1.837\pm 0.073$ \\
\hline
& $\beta_3^\omega = 11.326\pm 0.111$ &  $\beta_4^\omega = 0.3822\pm 0.0045$  \\
\hline
 & $\gamma_4^\omega = 33.268\pm 0.358$ & $\eta_2^\omega = -0.0277 \pm 0.0021$ \\
\hline
 & $\delta_3^\omega = 0.338 \pm 0.048$  &  $\delta_4^\omega = 7.800 \pm 0.721$  \\ 
\hline
\end{tabular}
\caption{} 
\label{tab:omega-c} 
\end{table} 


\subsection{Functional form of the $K^0_S$ channel at $Q_0^2 = 1$ GeV$^2$}
\label{sec:Akaon}

\subsubsection{up}
\vspace{-0.5cm}
\begin{equation}
\begin{split}
D_{1, K}^u (z, M_h; Q_0^2) &= 2 |\bm{R}| \,  
\exp \left[ P(\gamma_1^K, \gamma_2^K, \gamma_3^K, \gamma_4^K, 0; z) \right]  \\
&\times \Bigg[  \frac{2 (N_1^K)^2 \Delta M_h}{N} \, \mathrm{BW} (m_K, \Gamma_K; M_h)  + 
(\eta_1^K)^2\, \exp \left[ P(0, 1, \delta_1^K, \delta_2^K, 0; M_h) + \delta_3^K z M_h \right]   \Bigg]   \; , 
\label{eq:kaon-u}
\end{split}
\end{equation}
where 
\begin{equation}
N = \int_{0.49}^{0.51} dM_h \, 2 |\bm{R}| \, \mathrm{BW} (m_K, \Gamma_K; M_h) \; , 
\label{eq:Knorm}
\end{equation}
with $m_K = 0.498$ GeV, $\Gamma_K = 10^{-8}$ GeV, and $\Delta M_h = 0.02$ GeV, and with best-fit parameters
\begin{table}[h]
\begin{tabular}{|c||c|c|}
\hline
$K^0_S$ &  &  \\
\hline
\hline
$u$ & $N_1^K  = 0.191\pm 0.027$ &   \\
\hline
 & $\gamma_1^K = 0.210\pm 0.049$ & $\gamma_2^K = 5.243\pm 0.477$ \\ 
 \hline
 & $\gamma_3^K = -2.922\pm 0.795$ & $\gamma_4^K = -5.270\pm 0.680$  \\
\hline
 & $\delta_1^K =  2.384\pm 0.110 $  &  $\delta_2^K = -5.043 \pm 0.080 $  \\
 \hline
 & $\delta_3^K = 0.633 \pm 0.091 $  & $\eta_1^K = 0.0634 \pm 0.0089$  \\ 
\hline
\end{tabular}
\caption{} 
\label{tab:kaon-u} 
\end{table} 

\subsubsection{down}
\vspace{-0.5cm}
\begin{equation}
\begin{split}
D_{1, K}^d (z, M_h; Q_0^2) &= (N_2^K)^2 z^{\alpha_1^K}  \, D_{1, K}^u (z, M_h; Q_0^2)  \; , 
\label{eq:kaon-d}
\end{split}
\end{equation}

with best-fit parameters

\begin{table}[h]
\begin{tabular}{|c||c|c|}
\hline
$K^0_S$ &  &   \\
\hline
\hline
$d$ & $N_2^K  = 1.373\pm 0.028$ & $\alpha_1^K = 0.426\pm 0.037$  \\
\hline
\end{tabular}
\caption{} 
\label{tab:kaon-d} 
\end{table} 

\subsubsection{strange}

\begin{equation}
\begin{split}
D_{1, K}^s (z, M_h; Q_0^2) &= (N_3^K)^2 z^{\alpha_2^K}  \, D_{1, K}^u (z, M_h; Q_0^2)  \; , 
\label{eq:kaon-s}
\end{split}
\end{equation}

with best-fit parameters

\begin{table}[h]
\begin{tabular}{|c||c|c|}
\hline
$K^0_S$ &  &   \\
\hline
\hline
$s$ & $N_3^K  = 2.551\pm 0.039$ & $\alpha_2^K = 0.766\pm 0.028$  \\
\hline
\end{tabular}
\caption{} 
\label{tab:kaon-s} 
\end{table} 

\subsubsection{charm}
\begin{equation}
\begin{split}
D_{1, K}^c (z, M_h; Q_0^2) &=  2 |\bm{R}| \,  
\exp \left[ P(\gamma_5^K, \gamma_6^K, \gamma_7^K, \gamma_8^K, 0; z) \right]  \\
&\times \Bigg[  \frac{(N_4^K)^2 \Delta M_h}{N} \, \mathrm{BW} (m_K, \Gamma_K; M_h)  + 
(\eta_2^K)^2\, \exp \left[ P(0, 1, \delta_4^K, \delta_5^K, 0; M_h) + \delta_6^K z M_h \right]  \Bigg]  \; , 
\label{eq:kaon-c}
\end{split}
\end{equation}

with best-fit parameters

\begin{table}[h]
\begin{tabular}{|c||c|c|}
\hline
$K^0_S$ &  &   \\
\hline
\hline
$c$ & $N_4^K  = 0.596\pm 0.096$ &   \\
\hline
 & $\gamma_5^K = 0.435\pm 0.076$ & $\gamma_6^K = 1.987\pm 0.729$ \\
\hline
 & $\gamma_7^K = 3.624\pm 1.660$ & $\gamma_8^K = -11.641\pm 1.351$ \\
\hline
 & $\delta_4^K = 2.723 \pm 0.154 $  &  $\delta_5^K =  -5.122\pm 0.116 $  \\
\hline
 & $\delta_6^K = -0.180 \pm 0.130 $ &  $\eta_2^K =  0.109\pm  0.018$   \\ 
\hline
\end{tabular}
\caption{} 
\label{tab:kaon-c} 
\end{table} 

\endwidetext


\bibliographystyle{apsrevM}
\bibliography{mybiblio}


\end{document}